\documentclass[aps,twocolumn,floats,pre]{revtex4}
\usepackage{graphics,graphicx,epsfig}
\usepackage{amssymb,color}
\usepackage{epsf,epstopdf,wrapfig}

\newcommand{\beq}{\begin{equation}}
\newcommand{\eeq}{\end{equation}}
\newcommand{\beqn}{\begin{eqnarray}}
\newcommand{\eeqn}{\end{eqnarray}}

\begin{document}

\title{Optimizing information flow in small genetic networks. II:  Feed forward interactions}

\author{Aleksandra M. Walczak,\footnote{awalczak@princeton.edu}$^a$
Ga\v{s}per Tka\v{c}ik,\footnote{gtkacik@sas.upenn.edu}$^b$   and William Bialek\footnote{wbialek@princeton.edu}$^a$}

\affiliation{$^a$Joseph Henry Laboratories of Physics, Lewis--Sigler Institute for Integrative Genomics, and
Princeton Center for Theoretical Science,
Princeton University,
Princeton, New Jersey 08544\\
$^b$Department of Physics and Astronomy, University of Pennsylvania, Philadelphia, Pennsylvania 19104--6396}

\date{\today}

\begin{abstract}
Central to the functioning of a living cell is its ability to control the readout or expression of information encoded in the genome.  In many cases, a single transcription factor protein activates or represses the expression of many genes. As the concentration of the transcription factor varies, the target genes thus undergo correlated changes, and this redundancy limits the ability of the cell to transmit information about input signals.  We explore how interactions among the target genes can reduce this redundancy and optimize information transmission.  Our discussion builds on recent work [Tka\v{c}ik et al, {\em Phys Rev E} {\bf 80,} 031920 (2009)], and there are connections to much earlier work on the role of lateral inhibition in enhancing the efficiency of information transmission in neural circuits; for simplicity we consider here the case where the interactions have  a feed forward structure, with no loops.  Even with this limitation, the networks that optimize information transmission have a structure reminiscent of the networks found in real biological systems. 
\end{abstract}

\maketitle

\section{Introduction}

The genomes of even the smallest bacteria encode the structure of hundreds of proteins; for complicated organisms like us, the number of different proteins reaches into the tens of thousands \cite{alberts}.  During the course of its life, each cell has to control how many copies of each protein molecule are synthesized \cite{ptashne+gann_02}.  These decision about the expression of genetic information occur on many scales, from  (more or less) irreversible decisions during differentiation into different tissue types down to continuous modulations of the number of enzyme molecules in response to varying metabolic needs and resources \cite{wagner}.  As with many biological processes, the control of gene expression depends critically on the transmission of information.  Because the relevant information is represented inside the cell by signaling molecules at low concentrations, the irreducibly stochastic behavior of individual molecules sets a physical limit to information transmission \cite{bialek+setayeshgar_05,bialek+setayeshgar_08,tkacik+al_08a,tkacik+al_08b}.  A more precise discussion is motivated by the emergence of experiments which measure the noise in gene expression \cite{elowitz+al_02, ozbudak+al_02, blake+al_03, raser+oshea_04, rosenfeld+al_05, pedraza+oudenaarden_05,gregor+al_07b}.  In this paper, building on our previous work \cite{tkacik+al_09}, we explore how cells can structure their genetic control circuitry to optimize information transmission in the presence of these physical constraints, thus maximizing the control power that they can achieve while using only a limited number of molecules.

Although the problems of information transmission in genetic control are general, it is useful to have  a concrete example in mind.  In developing embryos, information about position---and hence, ultimately, fate in the developed organism---is encoded by spatially varying concentrations of `morphogen' molecules \cite{wolpert_69,lawrence_92}.  These molecules often are transcription factors, and positional information then is transmitted to the expression levels of the target genes \cite{lawrence_92,gurdon+al_01,volhard_08}.  In this scheme, maximizing information transmission maximizes the richness  of the spatial patterns which the embryo can construct.

In recent work we have considered the problem of maximizing information transmission from a single transcription factor to multiple target genes, in the case where the targets are non--interacting \cite{tkacik+al_09}.  Already this problem generates some structures that remind us of real genetic control networks.  But, when a single transcription factor controls many genes independently, the signals carried by those genes necessarily are redundant \cite{note_re_neurons}.  To fully optimize the information which can be carried by a given number of molecules, there must be interactions among the target genes which  reduce this redundancy.  Our goal in this paper is to derive the form of these interactions.  
We emphasize that this is an ambitious project:  we are trying to find the topology and parameters of a genetic network from first principles, constrained only by the limited concentration of all the relevant molecules.  To simplify our task, we start here with the case in which the network of interactions has no closed loops, and return to the full problem in subsequent papers.

\begin{figure}
\includegraphics[width =  \linewidth]{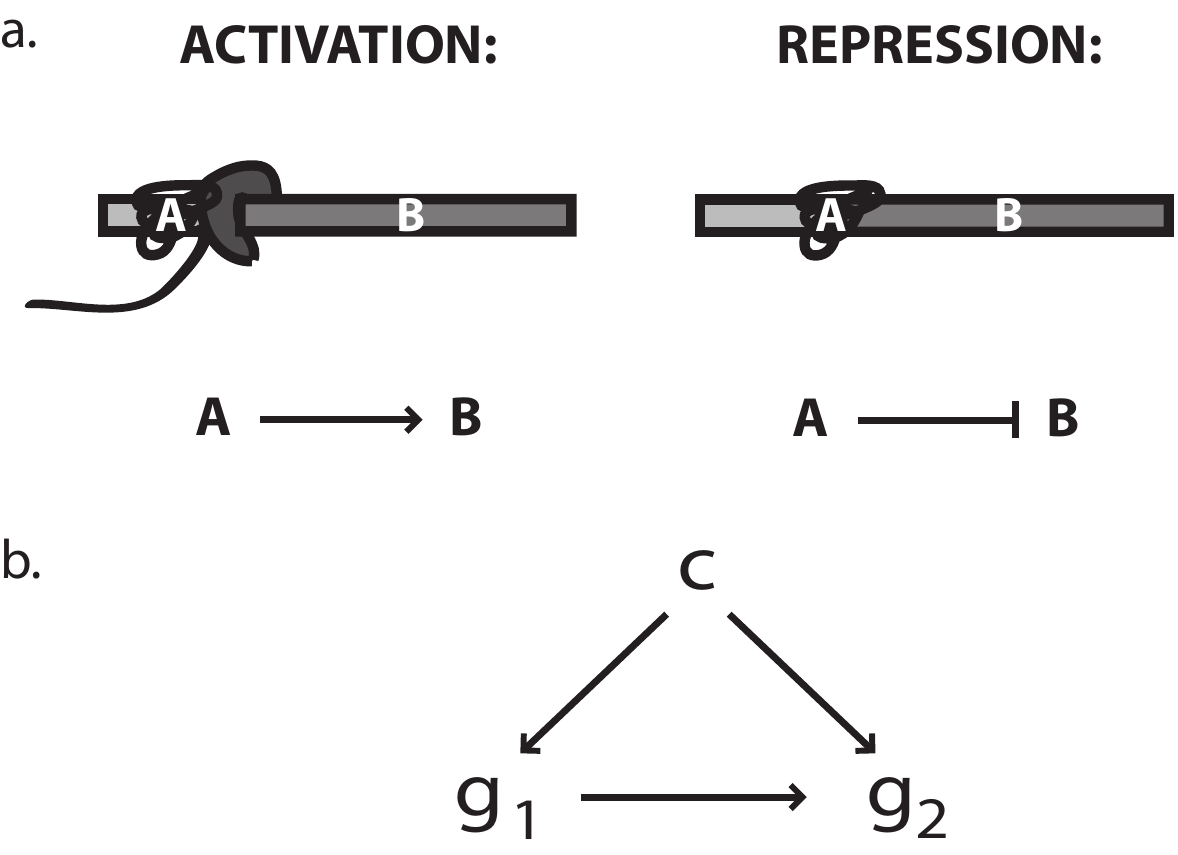}
\caption{Schematic models of transcriptional regulation.  (a.) Transcription factors proteins ($A$) can either activate (right) the expression of genes ($B$) by recruiting the RNA polymerase, or repress (left) the expression of genes by blocking the polymerase. Throughout the paper we will use the diagrammatic representation, in which an arrow depicts activation and a blunted arrow depicts repression. (b.) A feed forward network in which the input $c$ regulates the expression levels of $g_1$ and $g_2$, and $g_1$ also regulates $g_2$; the structure is feed forward because $g_2$ does not feedback on $g_1$. 
\label{scheme0}}
\end{figure}

When we search possible networks for the ones that optimize information transmission, it is convenient to break this search into two parts.  First, we enumerate network topologies, and then we search parameters within each topology.  It is conventional to encode in the network topology the sign of the interactions, so that changing the `shape of an arrow' from activation to repression is counted as a qualitatively different network (see Fig. \ref{scheme0}).   Even if we assume that each protein can act either as activator or repressor for all its targets \cite{dale},  with one transcription factor controlling $K$ interacting genes in a feed forward network, there are $2^K$ possible network topologies.   For each of these possibilities we find a single well defined optimum for all the continuously adjustable parameters in the system, and these optimal parameter values are determined only by the number of available molecules.  Thus, the structures of these networks really are derivable, quantitatively, from first principles.  For related approaches to information flow in biochemical and genetic networks, see Refs \cite{ziv+al_07,mugler+al_08,yu+al_08,tostvein+wolde_09,mehta+al_09}.

Among the possible network topologies, there are some in which interaction strengths are driven to zero by the optimization.  Among the nontrivial solutions, however, we find that the different topologies achieve very similar information capacities.  Thus, it seems that there is not a single optimal network, but rather an exponentially large number of nearly degenerate local optima.    All of these local optima are networks with competing interactions, so that a single target gene is activated and repressed by different inputs.  Phenomenologically, if we view the expression level of each target gene as a function of the input transcription factor concentration, these competing interactions lead to non--monotonic input/output relations.  It is this non--monotonicity which allows the collection of target genes to explore more fully the space of expression levels, thus enhancing the capacity to transmit information.  Although we consider only networks without feedback, these non--monotonic responses already are reminiscent of the patterns seen in real genetic networks.

\section{Setting up the problem}

The optimization problem that we are addressing in this paper is a generalization of the problem discussed in our previous work \cite{tkacik+al_09}.  To make this paper self--contained, however, we begin by reviewing  the general formulation of the problem.
We are interested in situations where a single transcription factor controls the expression level of several genes, labelled ${\rm i} = 1, 2, \cdots , K$.  The expression levels of these genes, $g_{\rm i}$, carry information about the concentration $c$ of the input transcription factor.   It is this information $I(c; \{ g_{\rm i}\})$ that we suggest is optimized by real networks.  

To derive the consequences of our optimization principle, we require several ingredients:  we need to describe the space of possible networks, we need to relate $I(c; \{g_{\rm i}\})$ to some calculable properties of these networks, we need to put these pieces together to form the objective function for our optimization problem, and finally we need to perform the optimization itself.    In the interests of proceeding analytically as far as possible down this path, before resorting to numerics, we adopt several approximations:  we consider the case where expression levels have reached steady state values consistent with the input transcription factor concentration, we assume that the noise in the system is small, and we restrict our attention to networks that have a feed forward structure.  We have discussed the first two approximations in our previous work \cite{tkacik+al_09}, and our focus here on feed forward networks is intended as a useful intermediate step between the case of non--interacting target genes (as in Ref \cite{tkacik+al_09}) and the case of arbitrary interactions, to which we will return in a subsequent paper.

\subsection{Describing the regulatory interactions}
\label{models}

The expression levels of genes are determined by a complex sequence of events, especially in eukaryotic cells where even the transcription of DNA into messenger RNA involves a complex of more than one hundred proteins \cite{Brivanlou02}.  As in much previous work, we abstract from this complexity to say that expression levels are set by a balance of synthesis and degradation reactions; in the simplest case only synthesis is regulated, and degradation is a first order process.  Then the deterministic kinetic equations that the govern the expression levels are
\begin{equation}
{{dG_{\rm i}}\over{dt}} = r_{\rm max} f_{\rm i} ( c; \{G_{\rm j}\}) - {1\over \tau} G_{\rm i} ,
\label{det1}
\end{equation}
where $\tau$ is the lifetime of the gene products against degradation, $r_{\rm max}$ is the maximum number of molecules per second that can be synthesized, and $f_{\rm i}(c; \{G_{\rm j}\})$ is the ``regulation function'' that describes how the synthesis rate is modulated by the other molecules in system.   The regulation functions, which are positive and range between zero and one, will be different for every gene, but we assume that the lifetimes and maximum synthesis rates are the same for all of the genes \cite{lifetime_note}.
In this formulation, $G_{\rm i}$ counts the number of molecules of the protein coded by gene $\rm i$, while it is conventional to define $c$ as a concentration.  Before we are done it will be useful to normalize these quantities.

The steady state expression levels, $\bar G_{\rm i} (c)$, are the solutions of the simultaneous nonlinear equations
\begin{equation}
\bar G_{\rm i} (c)= r_{\rm max}\tau f_{\rm i}(c ; \{\bar G_{\rm j}(c)\}) .
\end{equation}
In general there can be multiple solutions, corresponding to a network that has more than one stationary state; as noted above, we will confine our attention here to networks with a feed forward architecture, in which this can't happen. In Fig \ref{scheme0}b we present an example of a feed forward network, where an input $c$ regulates the expression of genes $g_1$ and $g_2$. The product proteins of $g_1$ also regulate gene $g_2$, but $g_2$ does not regulate $g_1$. Formally, the `no loops' condition defining feed forward networks means that there is some assignment of the labels $\rm i$ so that $f_{\rm i}$ only depends upon $G_{\rm j}$ for ${\rm j} < {\rm i}$, and hence, as will be important below, the matrix $\partial f_{\rm i}/\partial G_{\rm j}$ is lower triangular.

A deterministic system with continuous inputs and outputs can transmit an infinite amount of information.  In reality, information transmission is limited by the discreteness and randomness of the individual molecular events.  If this noise is small, it can be described by adding Langevin `forces' to the deterministic kinetic equations \cite{hasty+al_01a, hasty+al_01b, thattai+al_01,paulsson_04, kampen_07}. 
To the extent that synthesis and degradation rates can be written as depending on the instantaneous concentrations or expression levels, as we have done in Eq (\ref{det1}), then the system has no `hidden' memory, and  the Langevin terms which describe the noise in the system should have effectively zero correlation time.  We also will assume that the noise in the synthesis and degradation of different gene products are independent.  Then we can write
 \begin{eqnarray}
{{dG_{\rm i}}\over{dt}} &=& r_{\rm max} f_{\rm i} ( c; \{G_{\rm j}\}) - {1\over \tau} G_{\rm i} + \eta_{\rm i}(t)\\
\langle \eta_{\rm i} (t) \eta_{\rm j} (t') \rangle &=& {\cal N}_{\rm i} \delta_{\rm ij} \delta (t - t') ,
\label{langevin1}
\end{eqnarray}
where the ${\cal N}_{\rm i}$ are the spectral densities of the noise forces.

The spectral densities of noise have several components which add together.  First, if we take Eq (\ref{det1}) seriously, it describes synthesis and degradation as simple first order processes; the terms which appear in the deterministic equation as rates should then be interpreted as the mean rates of Poisson processes that describe the individual molecular transitions.  Then we have
\begin{equation}
{\cal N}_{\rm i}^{(1)} = r_{\rm max} f_{\rm i} ( c; \{G_{\rm j}\}) + {1\over \tau} G_{\rm i} .
\label{N1}
\end{equation}
We are interested in small fluctuations around the steady state, so in this expression we should set $G_{\rm i} = \bar G_{\rm i} (c)$.  Further, we can account for the possibility that synthesis is a multi--step process by adding  a `Fano factor' $F$ to the spectral density of synthesis noise, so we have
\begin{eqnarray}
{\cal N}_{\rm i}^{(1)} &=& F r_{\rm max} f_{\rm i} ( c; \{\bar G_{\rm j}(c)\}) + {1\over \tau} \bar G_{\rm i} (c) \\
&=& {{1+F}\over{\tau}} \bar G_{\rm i} (c) .
\label{N1a}
\end{eqnarray}
The Fano factor is a measure of the dispersion of the probability distribution,
the ratio of the variance to the mean; here the relevant random variable is the number of reaction events in a small window of time.  As an example, if synthesis occurs in bursts, then we expect   $F > 1$ \cite{thattai+al_01, mehta+al_08, iyer+09}. Burst of both proteins and mRNAs have now been experimentally observed in a number of bacterial and eukaryotic systems \cite{ozbudak+al_02,  golding+al_05, raj+06, choi+al_08}.  

A second source of noise comes from the diffusive arrival of transcription factor molecules at their targets.  To describe this, we should remember that the regulation of transcription depends on events that occur in a very small volume, of linear dimension $\ell$.  The relevant concentrations are those in this volume, or more colloquially at the point where the transcription factors bind; binding and unbinding events acts as localized sources and sinks for the diffusion equation which describes the spatiotemporal variations in concentration.  Intuitively, this coupling generates noise because the diffusion of the transcription factors into the relevant volume is a stochastic process, so that even with the global concentration fixed there are local concentration fluctuations \cite{berg+purcell_77}.
Analysis of this problem \cite{bialek+setayeshgar_05,bialek+setayeshgar_08,tkacik+bialek_09} shows that, so long as the time for diffusion through a distance $\ell$ is short, the net effect of the diffusive fluctuations is captured by a Langevin term
\begin{equation}
{\cal N}_{\rm i}^{(2)} = \left( r_{\rm max} {{\partial f_{\rm i}(c;\{G_{\rm j}\})}\over{\partial c}}\right)^2 {c\over{D\ell}} ,
\label{N2}
\end{equation}
where $D$ is the diffusion constant; again we should evaluate this at $G_{\rm j} = \bar G_{\rm j}(c)$ to describe the small fluctuations around the steady state.

The same arguments which apply to the input transcription factor also apply to each of the target gene products when they act as transcription factors.   If there are $G_{\rm i}$ molecules of gene product $\rm i$, then these proteins are present at concentration $G_{\rm i}/\Omega$, where $\Omega$ is the relevant volume \cite{volume},
and the analog of Eq (\ref{N2}) is
\begin{eqnarray}
{\cal N}_{\rm i}^{(3)} &=& \sum_{\rm k} \left( r_{\rm max} {{\partial f_{\rm i}(c;\{G_{\rm k}\})}\over{\partial (G_{\rm k}/\Omega) }}\right)^2 {{(G_{\rm k}/\Omega)}\over{D\ell}} \\
&=&
{\Omega\over{D\ell}} 
\sum_{\rm k} \left( r_{\rm max} {{\partial f_{\rm i}(c;\{G_{\rm k}\})}\over{\partial G_{\rm k} }}\right)^2 G_{\rm k}  .
\label{N3}
\end{eqnarray}

Finally, in putting all these terms together, it is convenient to measure expression levels in units of the maximum mean expression level, which is $\bar G_{\rm i} = r_{\rm max}\tau$ molecules; that is, we define $g_{\rm i} = G_{\rm i}/(r_{\rm max}\tau)$.  Then we have
\begin{widetext}
\begin{eqnarray}
\tau {{dg_{\rm i}}\over{dt}} &=& f_{\rm i} (c; \{g_{\rm j}\}) - g_{\rm i} + \xi_{\rm i} (t),\label{langevin_final}\\
\langle \xi_{\rm i}(t) \xi_{\rm j}(t')\rangle &=& \delta_{\rm ij} \delta(t-t') {1\over{r_{\rm max}^2}} \left[
{\cal N}_{\rm i}^{(1)} + {\cal N}_{\rm i}^{(2)} + {\cal N}_{\rm i}^{(3)}
\right] \\
&=&
\delta_{\rm ij} \delta(t-t')\tau
\left[
 {{1+F}\over{r_{\rm max}\tau}} \bar g_{\rm i}(c)
+ 
\left( {{\partial f_{\rm i}(c;\{g_{\rm k}\})}\over{\partial c}}\right)^2 {c\over{D\ell\tau}}
+ \sum_{\rm k} \left( {{\partial f_{\rm i}(c;\{g_{\rm k}\})}\over{\partial g_{\rm k} }}\right)^2 {{g_{\rm k}}\over{D\ell\tau (r_{\rm max}\tau/\Omega)}}
\right] {\Bigg |}_{\{g_{\rm k} = \bar g_{\rm k}(c)\}}.\nonumber\\
&&\label{noiseterms}
\end{eqnarray}
\end{widetext}

The parameter combinations which appear in these equations have  simple interpretations.  If $F =1$, the synthesis and degradation of each gene product molecule really is an independent Poisson process, and hence  $r_{\rm max}\tau$ is not just the maximum number of molecules that can be made (on average), but also the maximum number of {\em independent} molecules, $N_g$ in the notation of Ref \cite{tkacik+al_09}.  For more complex processes, where $F\neq 1$, we can write $N_g = 2r_{\rm max} \tau/(1+F)$.
Similarly, the combination $r_{\rm max}\tau/\Omega$ is a concentration,   the maximum (mean) concentration of the gene products.  Since these are acting as transcription factors, we will assume that this maximum concentration is also the maximum concentration of the input transcription factor, $c_{\rm max}$. 
As discussed previously \cite{tkacik+al_09}, there is a natural scale of concentration, $c_0 = N_g/(D\ell\tau)$, and if we use units in which $c_0 = 1$ we can write
\begin{widetext}
\begin{equation}
\langle \xi_{\rm i}(t) \xi_{\rm j}(t')\rangle 
=
\delta_{\rm ij} \delta(t-t'){\tau\over{N_g}}
\left[
 \bar g_{\rm i}(c)
+ 
\left( {{\partial f_{\rm i}(c;\{g_{\rm k}\})}\over{\partial c}}\right)^2 c 
+ {1\over{c_{\rm max}}} \sum_{\rm k} \left( {{\partial f_{\rm i}(c;\{g_{\rm k}\})}\over{\partial g_{\rm k} }}\right)^2 g_{\rm k}\right] {\Bigg |}_{\{g_{\rm k} = \bar g_{\rm k}(c)\}}.
\label{xi_corr_last}
\end{equation}
\end{widetext}
To complete our description of the system we need to specify the regulation functions $f_{\rm i} (c;\{g_{\rm j}\})$.  

\subsection{The regulation functions}

The central event in transcriptional regulation is the binding of the transcription factor(s) to their specific sites along the DNA.  These binding sites must be close enough to the start of the coding sequence of the target gene that binding can influence the initiation of transcription by the RNA polymerase and its associated proteins.  In the simplest geometric picture, which might actually be accurate in bacteria \cite{bintu+al_05a, bintu+al_05b, kuhlman+al_07, kinney+al_07}, we can imagine the molecular interactions being so strong that the normalized mean rate of transcription---what we write as the regulation function $f_{\rm i}(c;\{g_{\rm j}\})$---is essentially the probability that certain binding sites are occupied or unoccupied.  If  $n$ transcription factors bind cooperatively to nearby sites, and binding activates transcription, then we would expect
\beq
f(c)=\frac{c^n}{K^n+c^n},
\eeq
and if the binding represses transcription, so that the mean rate is proportional to the fraction of empty sites, we should have
\beq
f(c)=\frac{K^n}{K^n+c^n}.
\eeq
In each case, the constant $K$ measures the concentration for half--occupancy of the binding sites, and so $F = -k_B T \ln K$ can be interpreted as a binding energy for the transcription factor to its specific site along the genome.  In the biological literature these models are  called Hill functions, after Hill's classical discussion of oxygen binding to hemoglobin \cite{hill}.  While it is convenient to distinguish between activators and repressors, we note that, within the Hill function model, we can pass smoothly from one to the other by allowing $n$ to change sign.

\begin{figure}[b]
\centerline{\includegraphics[width = 0.85\linewidth]{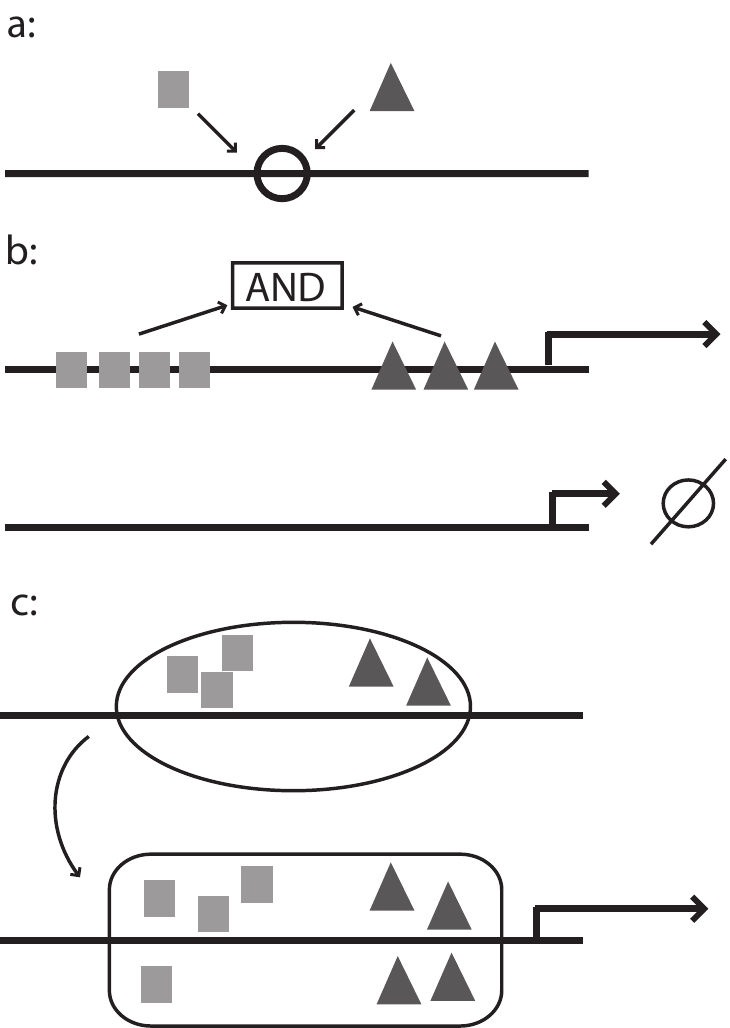}}
\caption{Regulation of transcription in the presence of two transcription factors.  (a.) Two types of transcription factors, squares and triangles, can bind to the regulatory region of the gene. This situation does not uniquely determine the form of regulation. (b.) The Hill regulatory model with the AND logic for the binding of the two types of transcription factors. In this case the both types of transcription factors must be bound to initiate transcription. The response of the gene is proportional to the product of the activation by each transcription factor separately. (c.) The MWC model for model for regulation. The transcriptional complex has two states, active and inactive, and binding of any type of transcription factor shifts the equilibrium between the two states. \label{fig0}}
\end{figure}

The Hill function model identifies the transcription of a gene as a logical function of binding site occupancy, so that the mean rate of transcription will be proportional to the probability of the sites being occupied (or, in the case of repressors, not occupied).  Thus there is a natural generalization when multiple transcription factors converge to control a single gene, as in Fig \ref{fig0}:  the rate of transcription should be proportional to logical functions on the occupancy of multiple binding sites.  Thus, if the input transcription factor and the gene $\rm j$ converge to activate gene $\rm i$, transcription could depend on the AND of the two binding events; if these are independent, then the regulation function becomes
\begin{equation}
f_{\rm i}(c, g_{\rm j}) = \frac{c^{n_{\rm i}}}{K_{\rm i}^{n_{\rm i}}+c^{n_{\rm i}}}
\frac{g_{\rm j}^{n_{\rm ij}}}{K_{\rm ij}^{n_{\rm ij}}+g_{\rm j}^{n_{\rm ij}}} .
\end{equation}
Alternatively, transcription could depend on the OR of the two binding site occupancies, in which case
\begin{equation}
f_{\rm i}(c, g_{\rm j}) = \frac{c^{n_{\rm i}}}{K_{\rm i}^{n_{\rm i}}+c^{n_{\rm i}}}
+
\frac{g_{\rm j}^{n_{\rm ij}}}{K_{\rm ij}^{n_{\rm ij}}+g_{\rm j}^{n_{\rm ij}}} ,
\end{equation}
and we can construct functions that interpolate or mix AND and OR.   In general we find that the pure AND functions transmit the most information. Thus  we will write
\begin{equation}
f_{\rm i}(c, \{g_{\rm j}\} ) = \frac{c^{n_{\rm i}}}{K_{\rm i}^{n_{\rm i}}+c^{n_{\rm i}}}
\prod_{{\rm j} < {\rm i}} \frac{g_{\rm j}^{n_{\rm ij}}}{K_{\rm ij}^{n_{\rm ij}}+g_{\rm j}^{n_{\rm ij}}} ,
\label{hill_model}
\end{equation}
where the restriction ${\rm j} < {\rm i}$ confines our attention to feed forward networks with no loops.  If $n_{\rm ij} >0$, then $K_{\rm ij} \rightarrow 0$ means that gene $\rm j$ does not regulate gene $\rm i$; if $n_{\rm ij} < 0$ then the condition for no interaction is $K_{\rm ij} \rightarrow \infty$.

An alternative model for regulation envisions a large transcriptional complex that can be in one of two states, and the rate of transcription is proportional to the population of the `active' state; transcription factors act by binding and shifting the equilibrium between the two states.  This is essentially the model proposed by Monod, Wyman and Changeaux for allosteric enzymes and cooperative binding \cite{MWC}.  The important idea is that, in each state, all effector molecules bind independently, but the binding energies are different in the two states.  By detailed balance, this difference in binding energy means that binding will shift the equilibrium between the two states.  In the case of one transcription factor at concentration $c$, with $n$ binding sites, the probability of being in the active state is given by:
\beq
f(c)=\frac{(1+c/K_{\rm on})^n}{L^n(1+c/K_{\rm off})^n+(1+c/K_{\rm on})^n},
\label{MWC_base}
\eeq
where $k_B T\ln L$ is the free energy different between the two states in the absence of any bound molecules.   If binding favors the active state, and  the gene has a very small probability to be ``on'' in the absence of input ($L\gg1$), then taking  $K_{\rm on}\ll K_{\rm off}$  reduces Eq (\ref{MWC_base}) to:
\begin{eqnarray}
f(c)&=&\frac{(1+c/K_{\rm on})^n}{L^n+(1+c/K_{\rm on})^n}\\
&=&
{1\over
{1+\exp\left[-n \ln\left({{1+c/K_{\rm on}}\over{ 1+c_{1/2}/K_{\rm on}}}\right)  \right]}
}
\label{MWC_use}
\end{eqnarray}
where we have introduced the constant $c_{1/2}$ to simplify notation, $L= (1+c_{1/2}/K_{\rm on})$, and $c_{1/2}$ is the input concentration for half maximal activation  \cite{unequalK}.  Note that activation and repression differ only in the sign of $n$. 

When more than one type of protein regulates expression (Fig \ref{fig0}c), 
each binding event makes its independent contribution to shifting the equilibrium between active and inactive states.  Thus, in the same limit of very unequal binding constants to the two states, we can write
\begin{eqnarray}
f_{\rm i}(c, \{g_{\rm j}\})&=& {1\over{1+\exp[-F_{\rm i}(c,\{g_{\rm j}\})]}}\\
F_{\rm i}(c,\{g_{\rm j}\}) &=& n_{\rm i} \ln\left({{1+c/K_{\rm i}}\over{ 1+c_{1/2}^{({\rm i})}/K_{\rm i}}}\right)  
\nonumber\\
&&\,\,\,\,\,
+ \sum_{{\rm j} < {\rm i}}n_{\rm ij} \ln\left({{1+g_{\rm j}/K_{\rm ij}}\over{ 1+\bar g_{\rm j} (c_{1/2}^{({\rm i})})/K_{\rm ij}}}\right)  .
\label{mwc_model}
\end{eqnarray}

For a single transcription factor, there is not much difference between the Hill and MWC models:  the classes of functions are very similar, and when we go through the optimization of information transmission, the optimal parameter values are such that the input/output relations are nearly indistinguishable.  But, as we shall see, when multiple transcription factors converge, the Hill and MWC models lead to significantly different behaviors.  This raises the possibility that the molecular details of the regulatory process are still `felt' at the more macroscopic, phenomenological level, but this is a huge set of questions that we leave aside for the moment, focusing just on these two models.


\subsection{Information transmission}

We are interested in connecting the properties of the networks described in the previous section to the information that the expression levels carry about the input transcription factor concentration.   This requires only a modest generalization of the discussion in Ref \cite{tkacik+al_09}, but, in the interest of clarity, we review the derivation.

In the limit that noise is small, the solution of the Langevin equations in Section \ref{models} has the form of Gaussian fluctuations around the steady state.  Formally, the probability distribution of expression levels given the concentration of the input transcription factor is
\begin{widetext}
\begin{equation}
P(\{g_{\rm i}\}| c)  = {1\over {(2\pi )^{K/2}}} \exp\left[ 
{1\over 2} \log \det( {\cal K}) - {1\over 2} \sum_{{\rm i, j}=1}^K
\left( g_{\rm i} - \bar g_{\rm i}(c)\right) {\cal K}_{\rm ij} 
\left( g_{\rm j} - \bar g_{\rm j}(c)\right)
\right] .
\end{equation}
\end{widetext}
The mean expression level of each gene, $\bar g_{\rm i}(c)$, defines an input/output relation for that gene, and the matrix $\cal K$ is the inverse covariance matrix of the fluctuations or noise in the expression levels at fixed input,
\begin{equation}
\langle \left( g_{\rm i} - \bar g_{\rm i}(c)\right) \left( g_{\rm j} - \bar g_{\rm j}(c)\right)\rangle = ({\cal K}^{-1})_{\rm ij} ;
\label{def_K}
\end{equation}
note that ${\cal K}$ is a function of $c$.  The fact that the expression levels ``carry information'' about the input transcription factor concentration means that it should be possible to estimate $c$ from knowledge of the $\{g_{\rm i}\}$.  From Bayes' rule, the distribution of concentrations consistent with some set of expression levels is given by
\begin{equation}
P(c|\{ g_{\rm i}\}) = {{P(\{g_{\rm i}\}| c) P(c)}\over{P(\{g_{\rm i}\})}} .
\end{equation}
In the limit that noise is small ($\cal K$ is large), this too is an approximately Gaussian distribution,
\begin{equation}
P(c|\{ g_{\rm i}\}) \approx {1\over\sqrt{2\pi \sigma_c^2}} \exp\left[ - {{\left( c-c_* (\{g_{\rm i}\})\right)^2}\over{2\sigma_c^2}}\right]
\label{gausscond},
\end{equation}
where $c_*(\{ g_{\rm i}\})$ is the most likely input given the output, and the effective variance is determined by
\begin{equation}
{1\over {\sigma_c^2 (\{g_{\rm i}\})}} = \sum_{{\rm i,j}=1}^K 
\left[
{{d\bar g_{\rm i}(c)}\over{dc}} {\cal K}_{\rm ij} {{d\bar g_{\rm i}(c)}\over{dc}}
\right]
{\Bigg |}_{c = c^*(\{g_{\rm i}\})} .
\end{equation}
In general it is difficult to find an explicit expression for $c_*(\{ g_{\rm i}\})$, but we will see that this is not crucial.
We can use these ingredients to calculate the amount of information, in bits, that the expression levels $\{g_{\rm i}\}$ provide about the input concentration $c$.

The mutual information between $c$ and $\{g_{\rm i}\}$ is defined, following Shannon \cite{shannon_48,cover+thomas_91}, as 
\begin{equation}
I(c;\{ g_{\rm i}\}) = \int dc\, \int d^K g \, P(c, \{g_{\rm i}\})\log_2 \left[
{{P(c, \{g_{\rm i}\})}\over{P(c)P( \{g_{\rm i}\})}}
\right] .
\end{equation}
We can rewrite this as a difference in entropies,
\begin{widetext}
\begin{equation}
I(c;\{ g_{\rm i}\}) = -\int dc\, P(c)\log_2 P(c) - \int d^K g \,P(\{g_{\rm i}\})
\left[  - \int dc\, P(c|\{ g_{\rm i}\}) \log_2 P(c|\{ g_{\rm i}\}) \right] ,
\end{equation}
\end{widetext}
where the first term is the entropy of the overall distribution of input concentrations, and the second term is the (mean) entropy of the distribution of inputs conditional on the output.  With the Gaussian approximation from Eq (\ref{gausscond}), we can evaluate
\begin{equation}
- \int dc\, P(c|\{ g_{\rm i}\}) \log_2 P(c|\{ g_{\rm i}\}) = {1\over 2}\log_2\left[
2\pi e \sigma_c^2(\{ g_{\rm i}\})
\right] .
\end{equation}
Further, when the noise is small, we can replace an average over expression levels by an average over inputs, setting the expression levels to their mean values along this path:
\begin{equation}
\int d^K g \,P(\{g_{\rm i}\}) [\cdots ] \approx
\int dc\, P(c) \prod_{\rm i} \delta \left( g_{\rm i} - \bar g_{\rm i}(c)\right) [\cdots ] .
\end{equation}
Putting these pieces together, we find an expression for the information in the low noise limit,
\begin{eqnarray}
I(c;\{ g_{\rm i}\}) &=& -\int dc\, P(c)\log_2 P(c)  \nonumber\\
&& \,\,\,\,\,\,\,\,\,\, - {1\over 2}\int dc\, P(c) 
\log_2\left[
2\pi e \sigma_c^2(\{ \bar g_{\rm i} (c)\})
\right] . \nonumber\\
&& \, \label{I_lownoise}
\end{eqnarray}

As Equation (\ref{I_lownoise}) makes clear, the information transmitted through the regulatory network depends on the input/output relations, on the noise level, {\em and} on the distribution of inputs $P(c)$.  The cell can optimize information transmission by adjusting this input distribution to match the characteristics of the network \cite{tkacik+al_08a,tkacik+al_08b}, but this matching is constrained by the cost of making the input molecules.  We can implement this constraint by trading bits against a cost function that counts the mean number of molecules, or more simply by insisting that $c$ is bounded by some maximum concentration $c_{\rm max}$.  In Ref \cite{tkacik+al_09} we have shown that these different ways of implementing the constraint given essentially identical results, so we use the $c_{\rm max}$ approach here.  We still need a constraint to force the normalization of $P(c)$, so we should maximize the functional
\beq
{\cal L}=  I(c, \{g_{\rm i}\})
- \lambda \int dc\, P(c) .
\eeq
The solution to this problem can be written, using the expressions for $I(c, \{g_{\rm i}\})$ from above, as
\begin{eqnarray}
P^*(c) &\propto& {1\over{\sigma_c(\{\bar g_{\rm i} (c)\})}}\\
&=& {1\over {Z}}
\left[{1\over{2\pi e}} \sum_{{\rm i,j}=1}^K 
{{d\bar g_{\rm i}(c)}\over{dc}} {\cal K}_{\rm ij} (c) {{d\bar g_{\rm j}(c)}\over{dc}}
\right]^{1/2}  ,
\label{Popt}
\end{eqnarray}
where the normalization constant
\begin{equation}
Z = \int_0^{c_{\rm max}} dc\, \left[{1\over{2\pi e}} \sum_{{\rm i,j}=1}^K 
{{d\bar g_{\rm i}(c)}\over{dc}} {\cal K}_{\rm ij} (c) {{d\bar g_{\rm j}(c)}\over{dc}}
\right]^{1/2}   .
\label{Z1_multiple}
\end{equation}
Finally, the information itself, evaluated with the optimal input distribution, is
\begin{equation}
I^* (c;\{g_{\rm i}\}) = \log_2 Z .
\end{equation}
This maximal mutual information still depends on all the parameters that describe the network, and our goal is to find the parameter values that maximize $I^*$.  To do this, it is useful to be a little more explicit about how the parameters enter into the computation of $Z$.

\subsection{Putting the pieces together}
\label{pieces}

To evaluate $Z$ in Eq (\ref{Z1_multiple}), we need to know the inverse covariance matrix ${\cal K}_{\rm ij}$ that describes the noise in gene expression.  This can be calculated from the Langevin equations in Section \ref{models}.  To do this, we recall that we are looking at the small noise limit, so we linearize the Langevin equations around the steady state $g_{\rm i} = \bar g_{\rm i}(c)$, and  then Fourier transform.  With $g_{\rm i}(t) = \bar g_{\rm i} + \delta g_{\rm i}(t)$, and
\begin{equation}
\delta g_{\rm i}(t) = \int {{d\omega}\over{2\pi}} e^{-i\omega t} \delta\tilde g_{\rm i}(\omega ) ,
\end{equation}
Eq (\ref{langevin_final}) becomes
\begin{equation}
\label{FTL}
\left[
\begin{array}{cccc}
1-i \omega \tau &0 & 0 &..\\
-\phi_{21} &1-i \omega \tau& 0&..\\
-\phi_{31}&-\phi_{32}&1-i \omega \tau&..\\
.. & ..&..&..
\end{array}
\right]
\left[
\begin{array}{c}
\delta\tilde g_1(\omega )  \\
\delta\tilde g_2(\omega )  \\
\delta\tilde g_3(\omega )  \\
..
\end{array}
\right]=
\left[
\begin{array}{c}
\tilde\xi_1(\omega )\\
\tilde\xi_2(\omega )\\
\tilde\xi_3(\omega )\\
..
\end{array}
\right],
\label{lin_langevin}
\end{equation}
where 
\begin{equation}
\phi_{\rm ij}=\frac{\partial f_{\rm i}(c, \{ g_{\rm k} \})}{\partial g_{\rm j}}
{\Bigg |}_{\{g_{\rm k} = \bar g_{\rm k}(c)\}}.
\end{equation}
In this expression we explicitly restrict ourselves to networks with a feedforward architecture, which (as noted above) means that we can assign labels $\rm i$ so that gene $\rm j$ regulates gene $\rm i$ only if ${\rm j} < {\rm i}$.  We recall from Eq (\ref{xi_corr_last}) that
\begin{equation}
\langle \xi_{\rm i}(t) \xi_{\rm j} (t')\rangle  =  \delta(t-t') N_{\rm ij} ,
\end{equation}
where the noise matrix is diagonal, 
\begin{widetext}
\begin{equation}
N_{\rm ij} = \delta_{\rm ij} {\tau\over{N_g}}
\left[
 \bar g_{\rm i}(c)
+ 
\left( {{\partial f_{\rm i}(c;\{g_{\rm k}\})}\over{\partial c}}\right)^2 c 
+ {1\over{c_{\rm max}}} \sum_{\rm k} \left( {{\partial f_{\rm i}(c;\{g_{\rm k}\})}\over{\partial g_{\rm k} }}\right)^2 g_{\rm k}\right] {\Bigg |}_{\{g_{\rm k} = \bar g_{\rm k}(c)\}}.
\end{equation}
\end{widetext}
The delta function in time means that the noise is white,
\begin{equation}
\langle \tilde\xi_{\rm i}(\omega ) \tilde\xi_{\rm j}^* (\omega ' )\rangle  =  2\pi\delta(\omega -\omega') N_{\rm ij} .
\end{equation}

Equations (\ref{FTL}) are of the form
\begin{equation}
\hat A (\omega) \cdot {\bf\delta \tilde g}(\omega)= \mathbf{\tilde\xi} (\omega) ,
\label{defA}
\end{equation}
where $\hat A(\omega )$ is the matrix describing the linearized dynamics,
$\bf{\delta \tilde{g}}(\omega) = \{ \delta \tilde g_1 (\omega ),\delta\tilde g_2 (\omega) , \cdots \}$, and similarly for  $\mathbf{\tilde\xi} (\omega)$.  Evidently
${\bf \delta \tilde{g}}(\omega)  = \hat A^{-1}(\omega ) \bf{\tilde\xi} (\omega) $.
We are interested in calculating the covariance matrix of the fluctuations $\bf\delta g$.  Since these are stationary, in the frequency domain we will have
\begin{equation}
\langle \delta\tilde g_{\rm i} (\omega ) \delta\tilde g_{\rm j}^* (\omega ' ) \rangle  = 2\pi\delta(\omega -\omega') S_{\rm ij}  (\omega ),
\end{equation}
and the equal time correlations that we want to calculate are related to the power spectrum through
\begin{equation}
\langle \delta g_{\rm i}\delta g_{\rm j} \rangle = \int {{d\omega}\over{2\pi}} S_{\rm ij} (\omega ).
\end{equation}
Following through the algebra, we have
\begin{widetext}
\begin{eqnarray}
\langle \delta\tilde g_{\rm i} (\omega ) \delta\tilde g_{\rm j}^* (\omega ' ) \rangle  
&=& [\hat A^{-1}(\omega )]_{\rm ik} [\hat A^{-1}(\omega' )]_{\rm jm}^* 
\langle \tilde\eta_{\rm k}(\omega ) \tilde\eta_{\rm m}^* (\omega ' )\rangle\\
&=& [\hat A^{-1}(\omega )]_{\rm ik} [\hat A^{-1}(\omega' )]_{\rm jm}^* N_{\rm km} 2\pi \delta (\omega - \omega')\\
\Rightarrow S_{\rm ij} (\omega ) &=& [\hat A^{-1}(\omega )]_{\rm ik} N_{\rm km} 
 [\hat A^{-1}( \omega )]_{\rm jm}^*
 = \left[ \hat A^{-1} (\omega) \hat N \left(\hat A^{-1} (\omega )\right)^\dag \right]_{\rm ij},
\end{eqnarray}
\end{widetext}
where $M^\dag$ denotes the Hermitian conjugate of the matrix $M$.  The covariance matrix, which is the inverse of ${\cal K}_{\rm ij}$ in Eq (\ref{def_K}), then takes the form
\begin{equation}
\left({\cal K}^{-1}\right)_{\rm ij} = \int {{d\omega}\over{2\pi}} \left[ \hat A^{-1} (\omega) \hat N \left(\hat A^{-1} (\omega )\right)^\dag \right]_{\rm ij} .
\label{Kfinal}
\end{equation}
To complete the calculation of $Z$ we need to do this frequency integral, and then an integral over the concentration $c$, as in Eq (\ref{Z1_multiple}).  Some details of these calculations are given in the Appendix.  Here we draw attention to two key points.

First, because the noise $\mathbf{\xi}$ is white---that is, $\hat N$ is independent of frequency---all of the structure in the frequency integral comes from the inverse of the matrix $\hat A (\omega)$.  Since the matrix elements of $\hat A$ are linear in frequency, this results in an integrand which is a rational function of $\omega$, so the integrals can all be done  by closing a contour in the complex $\omega$ plane.  Again, examples are in the Appendix. 

Second, and most importantly, when we finish computing $Z$ it depends on all the parameters we want to optimize---e.g., in the Hill description of the regulation functions, all of the half--maximal points $K_{\rm ij}$ and the cooperativities $n_{\rm ij}$---and in addition it depends on the maximal concentration of the input transcription factor, $c_{\rm max}$.  But there are no other parameters in the problem.  Thus, if we search for the maximum of $Z$, and hence the optimum information transmission, we will find the parameters of the network as a function of the available number of molecules, with no remaining arbitrariness.

\begin{figure*}
\includegraphics[width = 0.45\linewidth]{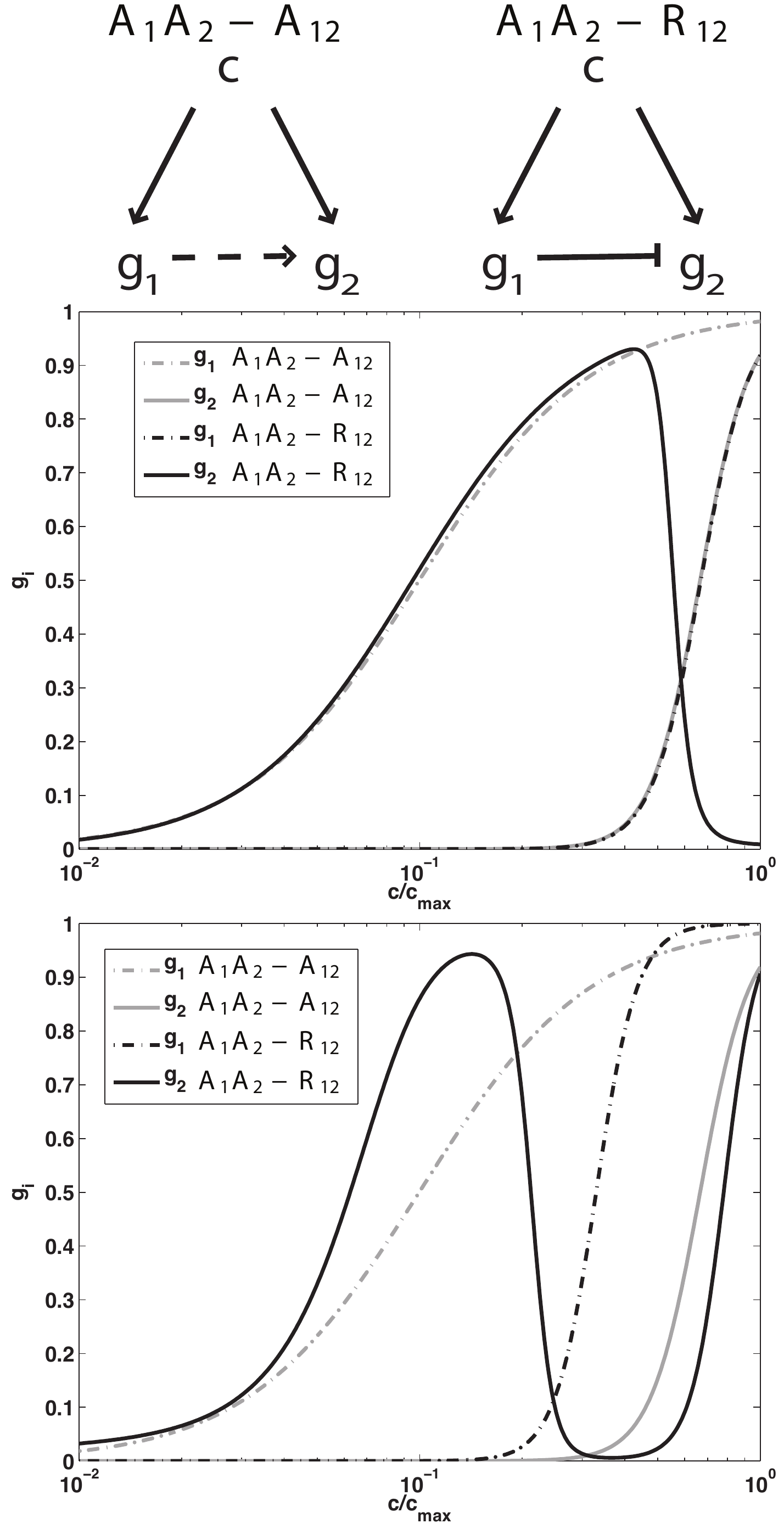}
\includegraphics[width = 0.45\linewidth]{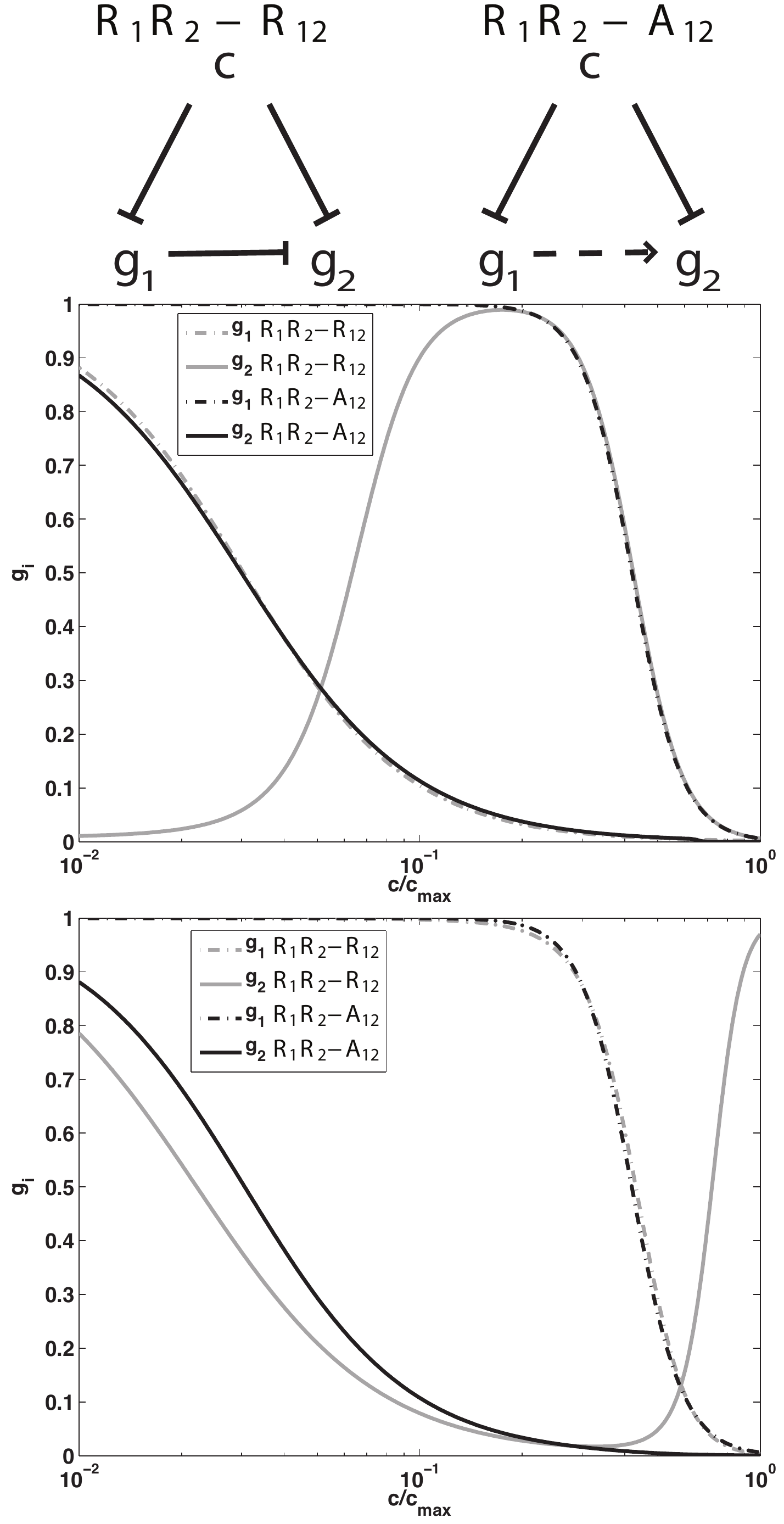}
\caption{Locally optimal networks for four topologies of the two gene feed forward networks, with $c_{\rm max}/c_0=10$.  Top panels are for the case of the Hill model, Eq (\ref{hill_model}), and bottom panels are for the MWC model, Eq (\ref{mwc_model}).  As explained in the text, the optimal parameters for the $A_1 A_2 - A_{12}$ and the $R_1 R_2 - A_{12}$ networks correspond to zero interaction between the target genes so these solutions also provide a reference for input/output relations in the optimal non--interacting networks.  Note that the non--interacting solutions are very similar for Hill and MWC, while the optimal interacting networks are qualitatively different.
\label{2genes}}
\end{figure*}

\section{Optimal networks and their parameters}

In previous work \cite{tkacik+al_09}, we considered the case of many noninteracting genes regulated by a single transcription factor. Optimizing information transmission in this systems leads to two qualitatively different families of solutions, depending on the available dynamic range $c_{\rm max}/c_0$.  At small $c_{\rm max}$,  the optimal strategy is for the different output genes to respond in the same way, allowing the system to make multiple independent ``measurements'' of the input concentration; in this case the target genes are completely redundant.
At large $c_{\rm max}$,   a tiling solution appears, in which the genes are turned on sequentially at different concentrations; the precise setting of the thresholds is determined by a compromise between minimizing the noise and maximizing the use of the dynamic range.  Even in this regime, however, the target genes are (partially) redundant.  Here we explore how the interactions between target genes can be tuned to reduce redundancy and increase the capacity of the system to transmit information.

\subsection{Two target genes}

We first  consider feed forward networks in which an input transcription factor at concentration $c$ regulates the expression of two genes, with normalized expression levels $g_1$ and $g_2$; in addition, the product proteins of gene 1 regulates the expression of gene 2.   Within this class of networks, we can calculate the information transmission following the outline above, and then search for optimal values of all the parameters.

We find that the general problem has many well defined local optima, corresponding to different topologies of the network, as shown in Fig \ref{2genes}.  With just two target genes and a single input, the constraint of considering feed forward networks means that the {\em locations} of the arrows in the graph describing the network are fixed, but it is conventional to distinguish different topologies based on the signs associated to the arrows, that is whether a transcription factor acts as an activator ($A$) or repressor ($R$), leaving us with multiple possibilities. For convenience we describe, for example, a network in which the input activates both target genes, while gene 1 represses gene 2, as $A_1 A_2 - R_{12}$.

Our intuition is that the role of interactions is to reduce redundancy, and this is borne out by the solutions associated to each topology.  In particular, if the input activates both targets, then having one target activate the other can't help to reduce redundancy, and correspondingly we find that in the $A_1 A_2 - A_{12}$ topology the interaction between the target genes is actually driven to zero by the optimization, and this is true whether we use the Hill model of regulation or the MWC model (top and bottom left panels of Fig \ref{2genes}).  The same intuition suggests that  the $R_1 R_2 - A_{12}$ topology should also be driven to zero interaction, and this too is what we find (right panels of Fig \ref{2genes}).   As an aside, these comparisons also verify that, in the non--interacting case, the Hill and MWC models achieve very similar input/output relations once we find the parameters that optimize information transmission.

\begin{figure}[b]
\includegraphics[width =  \linewidth]{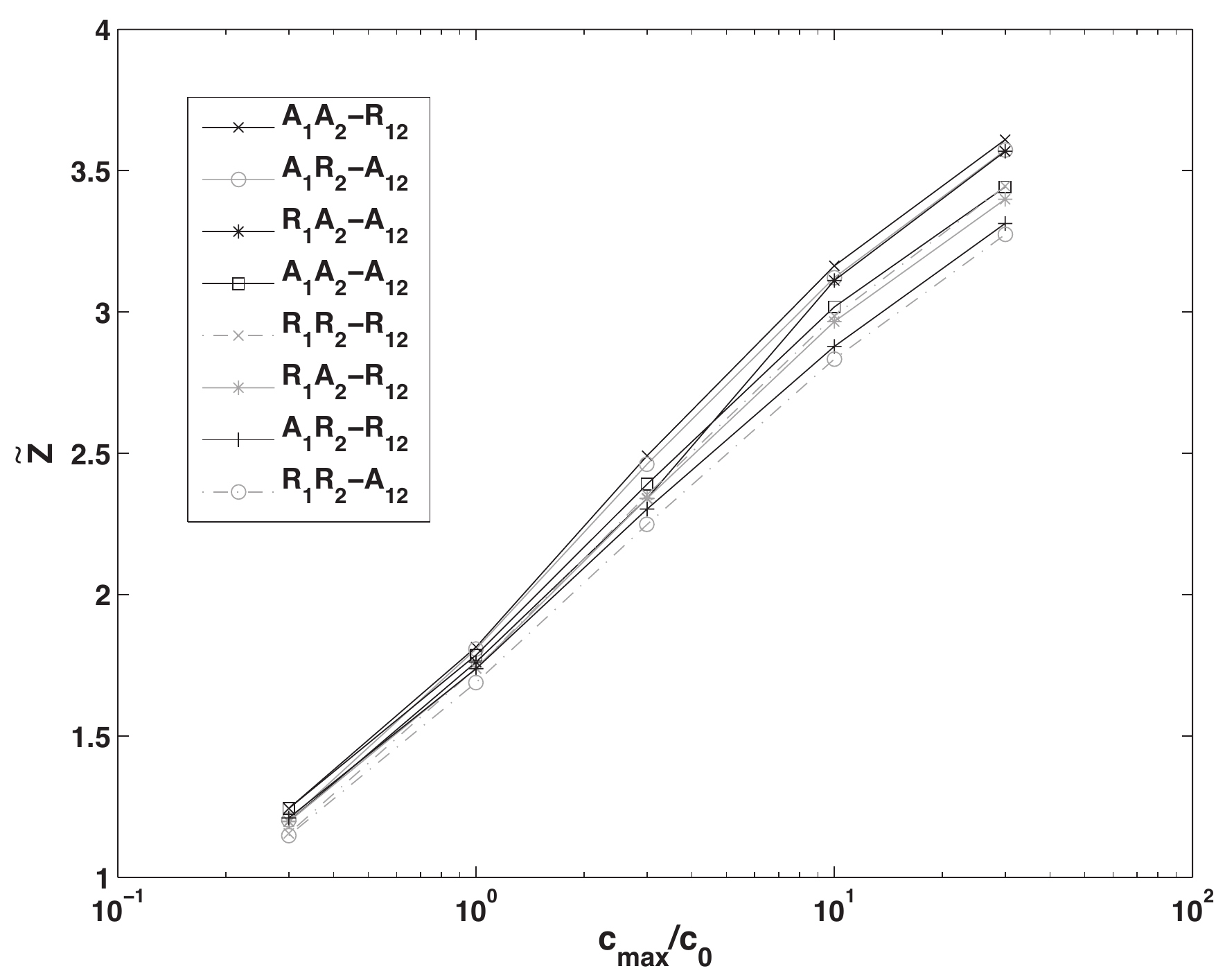}
\caption{Optimal information transmission in different network topologies, with Hill model regulation.  The information capacity of a network depends on $Z$, from Eq (\ref{Z1_multiple}), which in turn is proportional to the number of independent copies of the output molecules, $N_g$.  Here we plot $\tilde Z = 2\pi e Z /N_g$ as a function of the maximal concentration of input transcription factor, $c_{\rm max}/c_0$.
\label{cmax2FF}}
\end{figure}

\begin{figure}[t]
\includegraphics[width = 1\linewidth]{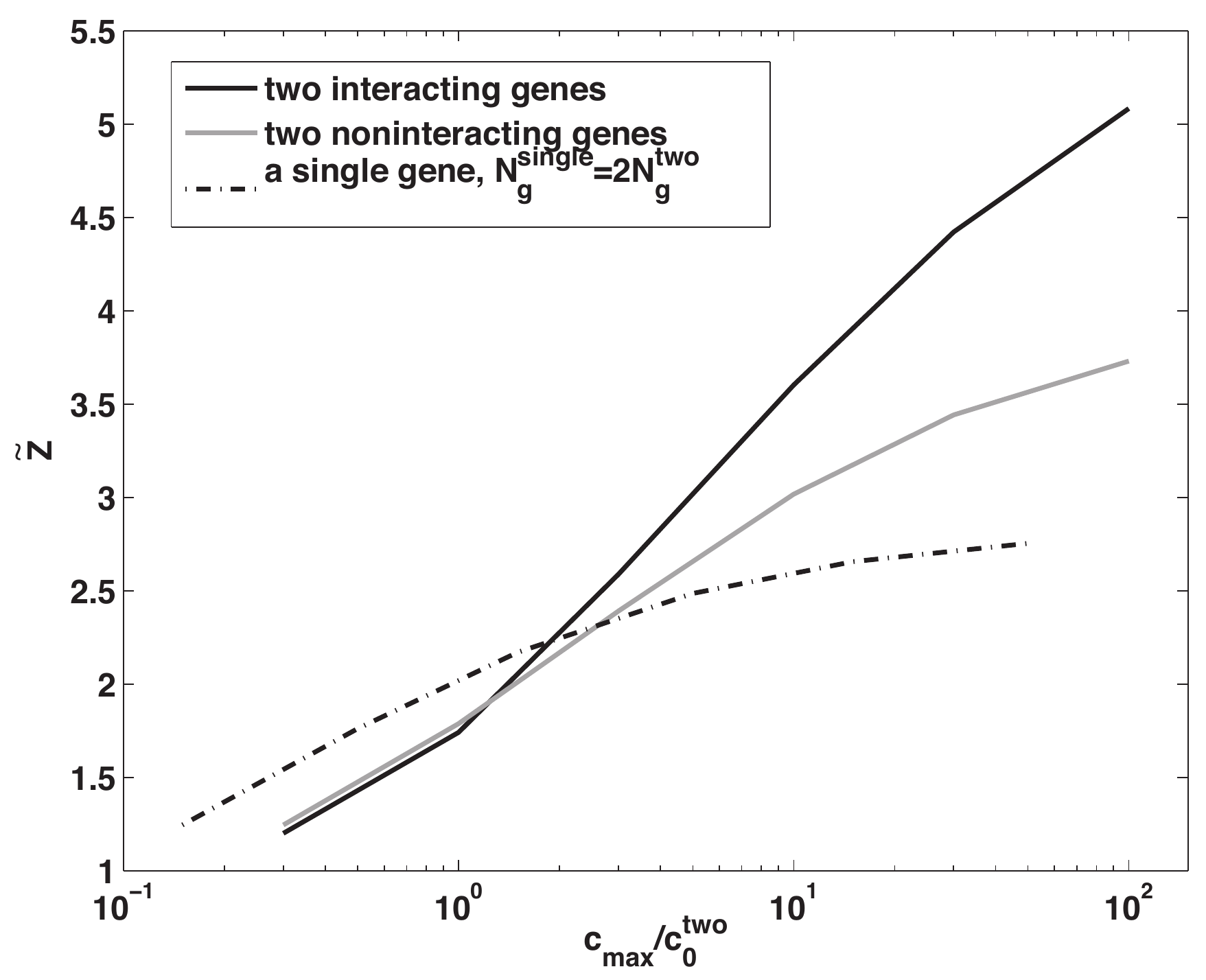}
\caption{One target or two? We show the dependence of $\tilde{Z}$ on the maximum concentration of input molecules, in the case of Hill regulation functions, comparing an interacting two gene network, a non--interacting two gene network, and a single output gene with twice the number of molecules.  Note that to make the comparison meaningful, we have to be careful about the definition of $c_0$, since this concentration scale includes a factor of $N_g$; here we normalize to $c_0$ for the two gene networks.
\label{FigNmax}}
\end{figure}

The networks which are optimized with non--trival interactions are the ones where one target gene represses the other, $A_1 A_2 - R_{12}$ and $R_1 A_R - R_{12}$.  For the Hill model, both of these networks are optimized when one of the target genes is expressed only in a finite band or stripe of input concentrations, so that if we monitor only the expression level $g_2$, we see a non--monotonic dependence on the input transcription factor.  If we think of the input concentrations as varying along one axis of a two dimensional space---as they do in the genetic networks controlling embryonic development \cite{wolpert_69,lawrence_92,gurdon+al_01,volhard_08}---then this non--monotonic behavior means that one of the genes will be expressed in a ``stripe'' through the middle of the spatial domain.  The other gene, which is itself a repressor, is expressed only at the extremes of the concentration range, either at high or low input concentrations depending on whether the input is an activator or repressor, respectively.  
As in the case of non--interacting target genes, all of this structure depends on the available dynamic range of inputs.  As we make $c_{\rm max}/c_0$ smaller, the optimal strength of the interactions becomes smaller, and the networks that optimize information transmission are more nearly non--interacting;  if $c_{\rm max}/c_0$ is sufficiently small, the optimal solution again is for all targets to be completely redundant, allowing the system to make multiple independent measurements of the same signal, rather than using the different targets to respond to different portions of the input dynamic range.

\begin{figure*}
\includegraphics[width = \linewidth]{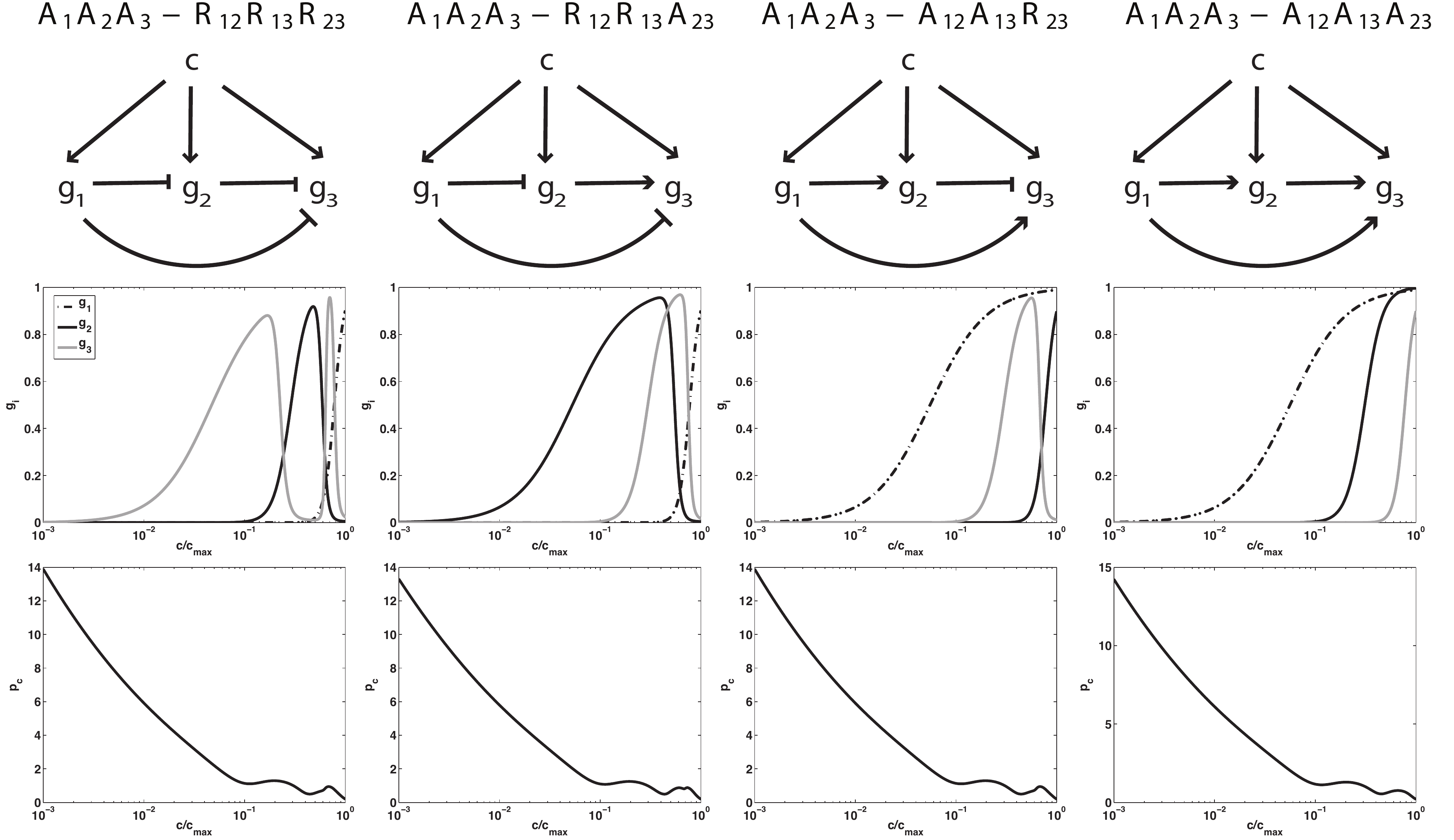}
\caption{Locally optimal networks with the three target genes, with the input acting as an activator.  Regulation is according to the Hill model, Eq (\ref{hill_model}), and we choose parameters to maximize information transmission with $c_{\rm max}/c_0=10$.   Top row shows  the network topologies, and the middle row shows the average expression levels of the different target genes, as in Fig \ref{2genes}.  The bottom panel shows the distribution of input concentrations that is matches the optimal networks, from Eq (\ref{Popt}).  
\label{3genesACT}}
\end{figure*}

We know from the analysis of the non--interacting system that sub--optimal settings of all the parameters really do incur large penalties, substantially reducing information transmission \cite{tkacik+al_09}.  This remains true for the interacting case, in that if we choose a topology but set the parameters arbitrarily, there is a significant loss of information.  On the other hand, once we choose parameters optimally, the different topologies have very similar information capacities, at least in the case of two target genes, as shown in Fig \ref{cmax2FF}.  To set a scale for these results on information transmission, we could ask whether there really is an advantage to having two target genes.  After all, noise is reduced by making more molecules, and with two targets the maximal number of output molecules will be increased.  A useful comparison, then, is with a system that has just one target gene, but can generate twice as many output molecules, and this is shown in Fig \ref{FigNmax}.  We see that, as the structure of the optimal networks collapses at low values of $c_{\rm max}/c_0$, so too does the advantage of having two separate target genes.  At very low concentrations of the input transcription factors, it really is better to have one target with twice the number of molecules, but as soon as there are enough molecules available to favor two target genes, the interactions between these genes become important in building the networks that optimize information flow.

One of the most interesting results of the optimization problem is that the Hill and MWC models, which are almost indistinguishable in the non--interacting case, are driven to qualitatively different solutions in the interacting case.  The basic idea of how interactions relieve redundancy  can be understood in a limit where each gene is either fully expressed  ($\bar g = 1$) or completely turned off ($\bar g = 0$).  For the non--interacting network, for example with the $A_1 A_2$ topology, the genes ``turn on'' in sequence, and so the system accesses states $00$, $01$, and $11$ in order as the input concentration is  increased.   In the interacting network $A_1 A_2 - R_{12}$, the optimal Hill model (cf Fig \ref{2genes}, top left panel) accesses state $00$, $01$, and then switches to $10$, not quite achieving the state $11$.  In contrast, the MWC model, once optimized, accesses all the states in sequence, $00 \rightarrow 01\rightarrow 10\rightarrow 11$, as the input concentration is increased, as we can see from the lower left panel of Fig \ref{2genes}; the situation is reversed in the case of $R_1 R_2 - R_{12}$.  We can understand how this happens, because in the MWC model each binding event shifts the equilibrium between the active and inactive states; thus there is a competition on gene 2 between the activating effect of the input and the repressing effect of gene 1, and this competition continues throughout the range of input concentration.  In the MWC model, by adjusting parameters, it is possible that the final increase in input concentration up to $c_{\rm max}$ overwhelms the repressive effect of gene 1, and indeed this is what happens at the parameter values that optimize information transmission.

The difference between the Hill and MWC models indicates that the relatively microscopic physics of protein--DNA interactions that underlies transcriptional regulation can be ``felt'' at a more macroscopic, phenomenological level. In this spirit, we should also ask if it matters that we assume each protein can act only as an activator or a repressor, but not both. It is easy to imagine the alternative in which, by proper placement of the binding sites, the transcription factor can activate one gene by recruiting the polymerase, but can repress other genes by occlusion. We have analyzed these mixed networks, and find that the optimal information transmission in these cases is intermediate between the ``pure'' networks considered thus far.  Thus, considering these additional topologies does not change the our global picture of the problem, except to open the possibility of yet more local optima.

\begin{figure*}
\includegraphics[width = \linewidth]{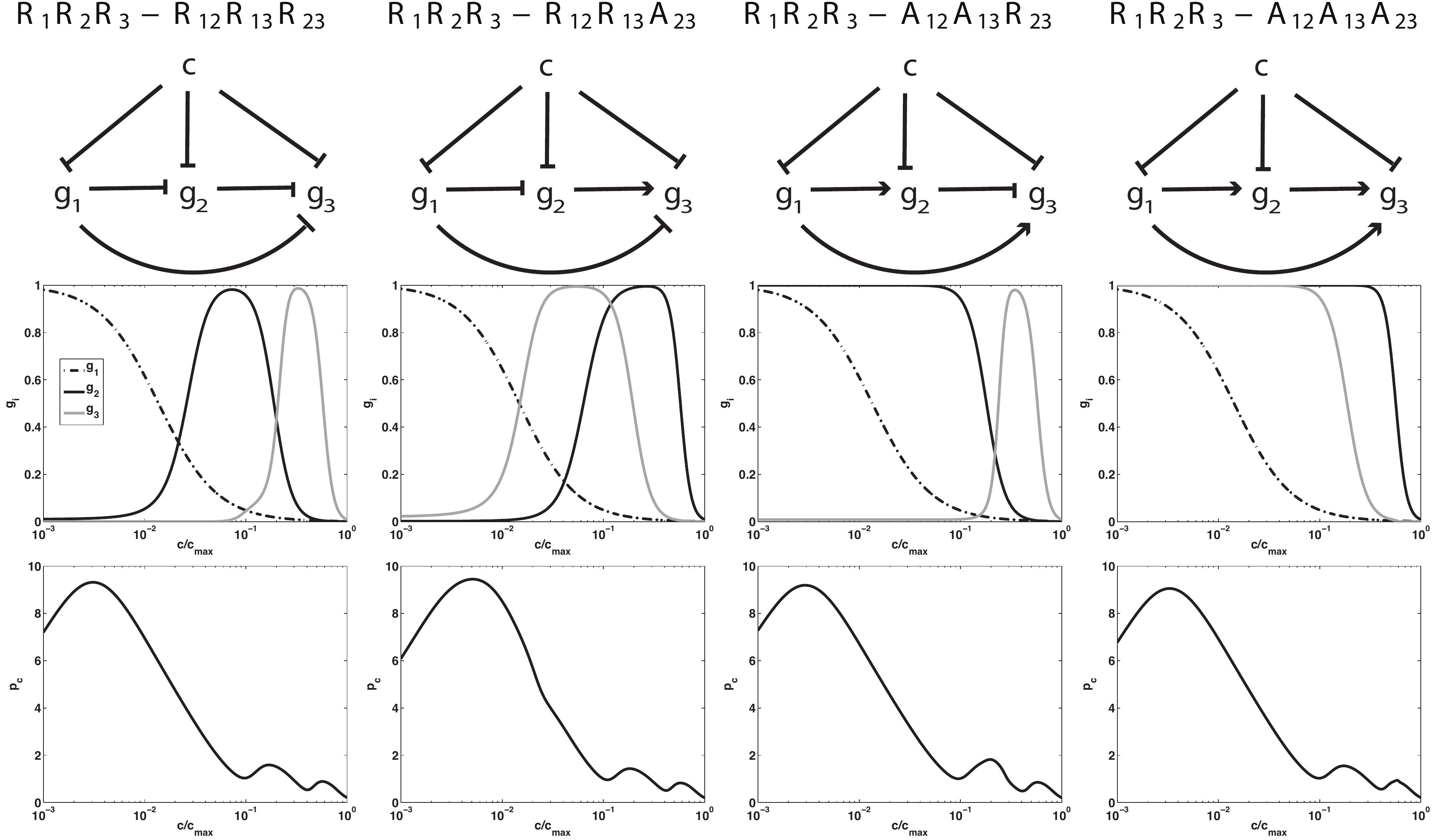}
\caption{Locally optimal networks with the three target genes, with the input acting as a repressor; all else as in Fig \ref{3genesACT}. \label{3genesREP}}
\end{figure*}

\subsection{Three target genes}

The analysis of networks with three target genes confirms and amplifies the lessons learned in the two gene case.  Because the space of possible networks is much larger, we once again restrict individual proteins to act only as activators or repressors, and give explicit results only for the case of the Hill function models of regulation, Eq (\ref{hill_model}); results are shown in Figs \ref{3genesACT} and \ref{3genesREP}.  As with two genes, having the target genes activate one another cannot reduce the redundancy that is generated in the non--interacting networks, and hence the $A_1A_2 A_3 -  A_{12}A_{13}A_{23}$ topology networks are driven back to the non--interacting $A_1 A_2 A_3$ when we solve the optimization problem, and similarly for  $R_1R_2 R_3 -  A_{12}A_{13}A_{23}$ Topologies that include some repressive interactions generate non--trivial solutions, with progressively richer structure the more repression we allow.  

The signature of the repressive interactions in the case of two targets was the emergence of non--monotonic dependencies of the expression levels on the input transcription factor concentration.  In the case of three targets,  the parameters which maximize information transmission again lead to non--monotonicity.  In the case where the input is an activator (Fig \ref{3genesACT}),   allowing for one repressive interaction  ($A_1 A_2 A_3 - A_{12}A_{13}R_{23}$) leads to one gene having a non--monotonic response.  If we allow two repressive interactions ($A_1 A_2 A_3 - R_{12}R_{13}A_{23}$) then  two genes have non--monotonic responses.  Finally, with the maximum of three repressive interactions ($A_1 A_2 A_3 - R_{12}R_{13}R_{23}$) one of the two 
non--monotonic responses becomes yet more complex, with two ``stripes'' of expression in different ranges of the input concentration.  In the case of an input repressor, we see a similar pattern (Fig \ref{3genesREP}).

\begin{figure}[b]
\includegraphics[width = 1\linewidth]{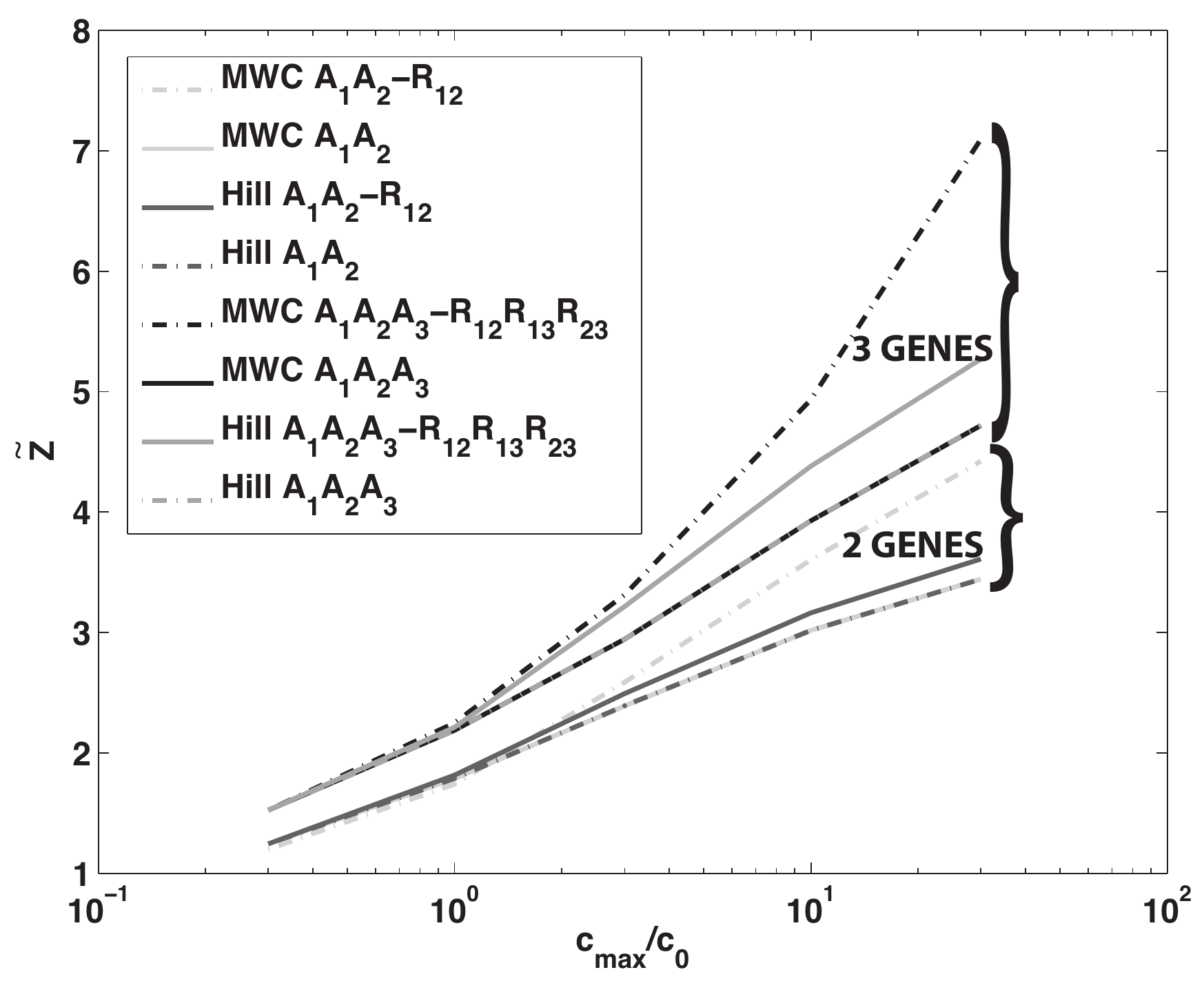}
\caption{Dependence of $\tilde{Z}$ on the maximum available protein concentrations  $c_{\rm{max}}/c_{\rm{0}}$ for 
two and three gene networks.  We show results for the non--interacting cases, and for the globally optimal topologies, using both the Hill and MWC models of regulation.
\label{HillMWC}}
\end{figure}

\begin{figure*}
\includegraphics[width = 0.49\linewidth]{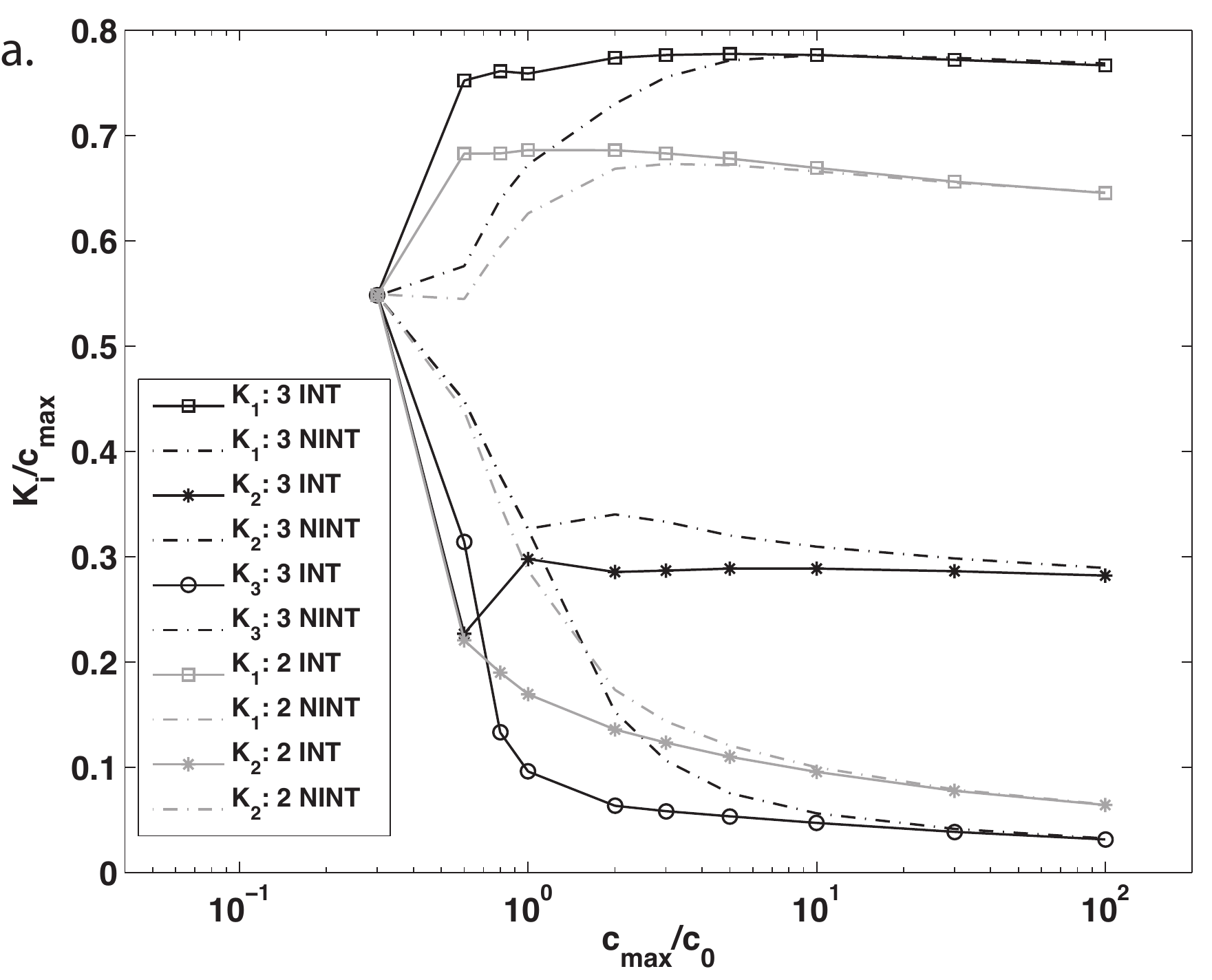}
\includegraphics[width = 0.49\linewidth]{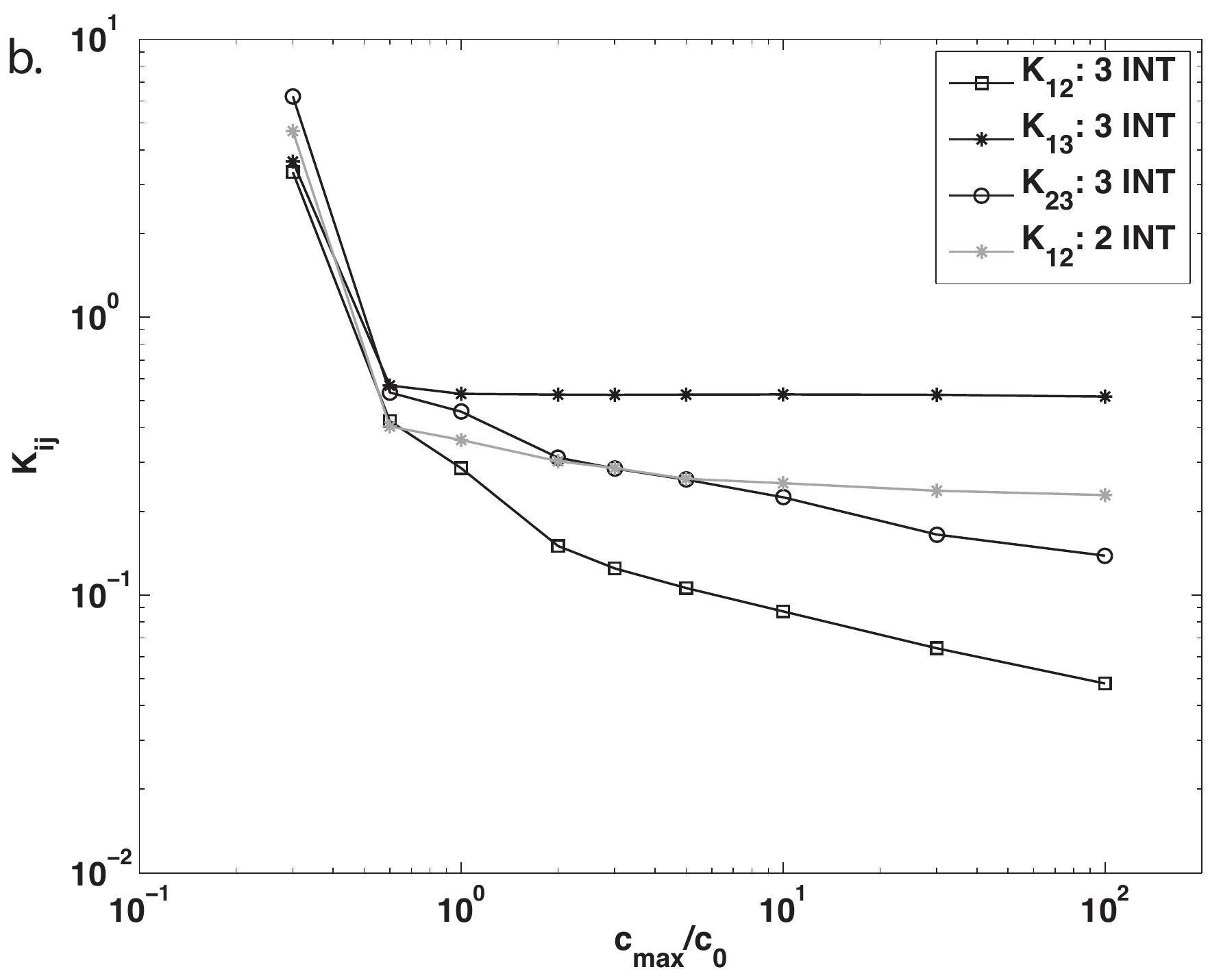}
\caption{Optimal parameters for Hill regulatory functions in the globally optimal topologies of  two ($A_1A_2-R_{12}$ grey lines) and three ($A_1A_2A_3-R_{12}R_{13}R_{23}$ black lines) gene networks compared to the parameters of the optimal noninteracting networks ($A_1A_2$, $A_1A_2A_3$). Parameters are presented as a function of  the maximum concentration of  available input proteins  $c_{\rm{max}}/c_{\rm{0}}$.  \label{PARAMSHill}}
\end{figure*}

A striking feature of the networks with two target genes was that local optima with different topologies were nearly degenerate.  This is less true for the case of three targets---the spread in capacities associated with the different topologies is larger, and the network $A_1 A_2 A_3 - R_{12}R_{13}R_{23}$ is the clear global optimum.  Corresponding to the larger spread in capacities, the enhancement of information transmission by interactions is larger, and in the difference in capacity between the Hill and MWC models also is larger in the three gene networks (Fig \ref{HillMWC}).  Including the possibility that individual transcription factors could act as both activators and repressors does not change these conclusions.

As a caveat, we note that we have assigned the same maximal concentration to all the transcription factors in the network.  This is equivalent to assuming that all these molecules come at the same cost to the organism.  One could imagine that there are significant differences between the molecules, and hence that one could be expressed at higher levels than the others for the same cost.  If this symmetry is broken explicitly, then the ordering of the local optima associated to different network topologies can be shifted.  While it is satisfying to identify a global optimum, it probably thus still makes sense to think of the problem as having many local optima that might all be relevant in different biological contexts.  An interesting feature of these different optima is that, despite the changing structure of the input/output relations for the target genes, the distribution of input concentrations that optimizes information transmission [cf Eq (\ref{Popt})] is almost the same for all the cases where the input is an activator (bottom panels of Fig \ref{3genesACT}) and again for all the cases where the input is a repressor (bottom panels of Fig \ref{3genesREP}).    Although maximizing information flow requires matching of the input distribution to the characteristics of the regulatory network \cite{tkacik+al_08a,tkacik+al_08b}, it seems that once the parameters of the network itself have been optimized, the solution to the matching problem has a more universal structure, with almost the same distribution providing the best match to several different local optima.

Finally, we consider the parameter values in these optimal networks.  For simplicity we focus on the globally optimal solutions with Hill model regulation functions for two ($A_1A_2-R_{12}$) and three ($A_1A_2-R_{12}R_{13} R_{23}$)  target genes, plotting in Fig \ref{PARAMSHill} the evolution of the different constants $K_{\rm i}$ and $K_{\rm ij}$ as a function of the maximal input concentration $c_{\rm max}/c_0$.  As noted above, the optimal  the interacting networks collapse onto the non--interacting case at small $c_{\rm max}$; for repressors, interactions become negligible if $K_{\rm ij}$ becomes large, and this is what we see at small $c_{\rm max}$.
At the opposite extreme, as $c_{\rm max}/c_0$ becomes large, the optimal $K_{\rm i}$ become nearly constant fractions of $c_{\rm max}$, while the interactions become stronger and stronger even up to $c_{\rm max}/c_0 \sim 10^2$.  Put another way, the concentrations at which the target genes are turned on by the input become distributed at fixed fractions of the available dynamic range, while the points at which the targets are turned off by their mutual repression evolves as a function of the maximal input concentration.  Further, in the interacting networks, the difference between two and three genes is not simply an extra target, but a readjustment of the interactions to make optimal use of this extra output channel.

\section{Discussion}

The expression levels of target genes provide information about the concentration of the transcription factors which provide input to the regulatory network.  Our goal here has been to understand how the structure of the network can be adjusted to optimize this information transmission.    Earlier work \cite{tkacik+al_09} addressed this problem in the limit where the target genes do not interact, so the input is simply ``broadcast'' to multiple targets; here we considered the role of interactions within a network of target genes, limiting ourselves, for simplicity, to feed forward structures.  We have found that there are an exponentially large number of locally optimal networks, each with parameters tuned to a well defined point that depends only on the maximal number of available molecules.  These optimal networks have two components.  As in the non--interacting case, the affinities of the input transcription factor for its multiple targets must be adjusted to balance the need to use the full dynamic range of inputs against the need to avoid noise at low input concentrations and output expression levels.  Unique to the interacting case, mutual repression among the target genes is tuned to reduce the redundancy among the network's outputs.

The problem of redundancy reduction has a long history in the context of neural coding, dating back to the first decade after Shannon's original work \cite{attneave_54,barlow_59,barlow_61}.  As first emphasized for the retina, nearby neurons often receive correlated signals; since the dynamic range of neural outputs is limited, this redundancy compromises the ability of the system to transmit information.  The solution to this problem is lateral inhibition---neighboring neurons that receive correlated inputs inhibit one another, so that each neuron transmits something which approximates the difference between its input and that of its neighbors.  Lateral inhibition, sometimes mediated through a more complex network, is a common motif in neural circuitry, and in the retina it is expressed by the ``center--surround'' organization of the individual neurons' receptive fields.  A more careful analysis shows that there is a tradeoff between redundancy reduction and noise reduction, so that the extent of inhibitory surrounds should be reduced, even to zero, as background light intensity is reduced and photon shot noise is increased, and this is in semi--quantitative agreement with experiment \cite{atick+redlich_90,hateren_92}.
The relation of these ideas in neural coding to our present discussion of genetic networks should be clear.  Although there are many questions about how these ideas connect more quantitatively to experiment, it is attractive to think that similar physical principles, and even analogous implementations of these principles, might be relevant across such a wide range of biological organization.

There are relatively few transcriptional regulatory networks that have been characterized to the point of comparing quantitatively with the sorts of models we have explored here.  In contrast, there is a growing literature on the qualitative, topological structure of these networks.  In particular, a number of groups have focused on the local ``motif'' structure of large networks, searching for patterns of interaction that are over--represented relative to some randomized ensemble of networks  \cite{shenorr+al_02, alon_07}.   One such motif is the feed forward loop, which essentially captures the whole set of feed forward networks with two target genes that we have considered here.  In the language of Fig \ref{scheme0}b, much of the discussion about the importance of this motif has centered on the nature of the transformation from the input concentration $c$ to the expression level of gene 2.  By choosing coherent (in our notation, $A_1A_2 - A_{12}$) or incoherent ($A_1A_2 - R_{12}$) loops, the system can achieve different temporal dynamics as well as non--monotonic input/output relations \cite{mangan+al_03a,mangan+al_03b,ishihara_05,wall+al_05,mangan+al_05,kaplan+al_08}.  Since we have considered system in steady state, it is this last point which is significant for our discussion.  Our results comparing Hill and MWC models of regulation emphasize, however, that it is not just the topology of the network which is important for the structure of these input/output relations.

One of the best opportunities for quantitative comparison with experiment is in the genetic networks controlling the early events of embryonic development.  In these systems, the primary morphogen molecules are sometimes transcription factors, so that we can literally see the variations in input concentration laid out in space \cite{lawrence_92,gurdon+al_01,volhard_08}.  Along the anterior--posterior axis of the {\em Drosophila} embryo, for example, the maternally provided morphogen Bicoid varies in concentration almost exponentially, so that the long axis of the embryo is essentially a logarithmic concentration axis \cite{houchmandzadeh+al_02,gregor+al_07a}.  Bicoid is a transcription factor that provides input to the network of gap genes, {\em hunchback}, {\em kr\"uppel}, {\em giant} and {\em knirps}.  Although Bicoid is an activator, these genes exhibit non--monotonic expression levels, forming stripes along the length of the embryo, and these non--monotonicities are mediated by mutual inhibitory interactions \cite{jackle+al_86,rivera-pomar+jackle_96,sanchez+thieffry_01,levine+al_05,yu+small_08}.  Qualitatively, the structure of the optimal networks that we derive here (especially the globally optimal $A_1A_2 A_3- R_{12}R_{13}R_{23}$) resembles the gap gene network, and the expression profiles along the $\log c$ axis are reminiscent of the spatial profiles of gap expression.  Generalizations of the experiments in Ref \cite{gregor+al_07b} should make it possible to map the input/output relations more quantitatively, and perhaps even to measure the interactions directly by detecting the predicted correlations among the noise in the expression levels of different gap genes.  To make truly quantitative comparisons, however, we need to solve our optimization problem allowing for networks with feedback.

\section*{APPENDIX}

This Appendix collects some technical details in the calculation of the inverse covariance matrix $\cal K$, following the framework established in Section
\ref{pieces}. We start by considering the case of one input transcription factor at concentration $c$ controlling the expression of two output genes, with normalized expression levels $g_1$ and $g_2$.  In this case the Langevin equations of motions describing the  network are:
\begin{eqnarray}
\tau {{dg_{\rm 1}}\over{dt}} &=& f_{\rm 1} (c) - g_{\rm 1} + \xi_{\rm 1} (t),\label{langevin_gene21}\\
\tau {{dg_{\rm 2}}\over{dt}} &=& f_{\rm 2} (c, g_{\rm 1}) - g_{\rm 2} + \xi_{\rm 2} (t)\label{langevin_gene21}.
\end{eqnarray}
We linearize the equations of motion around the steady state solutions $\{\bar{g}_{\rm 1}, \bar{g}_{\rm 2} \}$ and Fourier transform to obtain Eq (\ref{defA}), where the matrix $\hat A$ is explicitly given by:
\begin{equation}
\hat A (\omega ) = 
\left[
\begin{array}{cc}
1-i \omega \tau &0 \\
-\phi_{21} &1-i \omega \tau 
\end{array}
\right],
\end{equation}
and hence
\begin{equation}
\hat A^{-1} (\omega ) = 
{1\over{(1- i\omega \tau)^2}}
\left[
\begin{array}{cc}
1-i \omega \tau &0 \\
\phi_{21} &1-i \omega \tau 
\end{array}
\right] .
\end{equation}
The covariance matrix then can found, from Eq (\ref{Kfinal}):
\begin{widetext}
\begin{eqnarray}
 {\cal K}^{-1} &=& \int {{d\omega}\over{2\pi}} \left[ \hat A^{-1} (\omega) \hat N \left(\hat A^{-1} (\omega )\right)^\dag \right] \nonumber\\
&=& \int {{d\omega}\over{2\pi}} 
{1\over{(1- i\omega \tau)^2}}
\left[
\begin{array}{cc}
1-i \omega \tau &0 \\
\phi_{21} &1-i \omega \tau 
\end{array}
\right] 
\left[
\begin{array}{cc}
N_{11} &0 \\
0&N_{22} 
\end{array}
\right] 
{1\over{(1+ i\omega \tau)^2}}
\left[
\begin{array}{cc}
1+ i \omega \tau & \phi_{21} \\
0 &1+i \omega \tau 
\end{array}
\right] \\
&=& 
\int {{d\omega}\over{2\pi}}
{1\over{[1+(\omega\tau)^2]^2}}
\left[
\begin{array}{cc}
N_{11}[1+(\omega\tau)^2] & \phi_{21}N_{11}(1-i\omega\tau) \\
\phi_{21}N_{11}(1+ i\omega\tau) &N_{22}[1+(\omega\tau)^2]  + \phi_{21}N_{11}(1+ i\omega\tau) 
\end{array}
\right]\\
&=& 
\int {{d\omega}\over{2\pi}}
{1\over{[1+(\omega\tau)^2]^2}}
\left[
\begin{array}{cc}
N_{11}[1+(\omega\tau)^2] & \phi_{21}N_{11}  \\
\phi_{21}N_{11} &N_{22}[1+(\omega\tau)^2]  + \phi_{21}N_{11}  
\end{array}
\right] .
\end{eqnarray}
\end{widetext}
The elements of $N_{\rm ij}$ are given explicitly by: 
\begin{eqnarray}
N_{11}&=&{\tau\over{N_g} } \left[\bar g_{\rm 1}(c)+\left( {{\partial \bar{g}_{\rm 1}(c)}\over{\partial c}}\right)^2 c \right] \\
N_{22}&=&{\tau\over{N_g}} \left[\bar g_{\rm 2}(c)+\left( {{\partial \bar{g}_{\rm 1}(c)}\over{\partial c}}\right)^2 c
+{ \bar{g}_{\rm 1}\over{c_{\rm max}}}\phi_{21}^2 \right] ,
\end{eqnarray}
where we have chosen the maximum concentrations for all types of proteins to be equal to $c_{\rm max}$, and we have set $c_0=1$, as described in the text. To finish the calculation we need the integrals
\begin{eqnarray}
\int {{d\omega}\over{2\pi}}
{1\over{[1+(\omega\tau)^2]}} &=&
{1\over {2\tau}} ,\\ \nonumber\\
\int {{d\omega}\over{2\pi}}
{1\over{[1+(\omega\tau)^2]^2}} &=& {1\over {4\tau}} .
\end{eqnarray}
Then
\begin{widetext}
\begin{eqnarray}
{\cal K}^{-1} &=& {1\over {2\tau}} \left[
\begin{array}{cc}
N_{11} & \phi_{21}N_{11}/2  \\
\phi_{21}N_{11}/2 &N_{22}  + \phi_{21}N_{11} /2
\end{array}
\right] \\
{\cal K} &=&
{{2\tau}\over{ N_{11}(N_{22} + \phi_{21}N_{11}/2) - N_{11}^2 \phi_{21}^2/4}}
\left[
\begin{array}{cc}
N_{22}  + \phi_{21}N_{11} /2 & -\phi_{21}N_{11} /2  \\
- \phi_{21}N_{11}/2 & N_{11}
\end{array}
\right]
\end{eqnarray}
\end{widetext}
Lastly, we substitute the obtained covaiance matrix into the expression for the normalization constant $Z$, as obtained in Eq (\ref{Z1_multiple}):
\begin{equation}
Z = \int_0^{c_{\rm max}} dc\, \left[{1\over{2\pi e}} \sum_{{\rm i,j}=1}^2 
{{d\bar g_{\rm i}(c)}\over{dc}} {\cal K}_{\rm ij} (c) {{d\bar g_{\rm j}(c)}\over{dc}}
\right]^{1/2}   .
\label{Z1_multipleag}
\end{equation}
Since the optimal information is proportional to $Z$, we now can optimize $Z$ over the parameters of the regulation function. It is now clear, having set the concentration variables in terms of the natural scale $c_0=1$, that the only  parameter in the problem is $c_{\rm max}$; all other parameters will be determined by maximizing $Z$.

In the case of a three gene network, the calculation is analogous to the two gene case. The matrix $\hat A$ is explicitly given by:
\begin{equation}
\hat A (\omega ) = 
\left[
\begin{array}{ccc}
1-i \omega \tau &0&0 \\
-\phi_{21} &1-i \omega \tau &0\\
-\phi_{31} &-\phi_{21}&1-i \omega \tau
\end{array}
\right],
\end{equation}.
and the elements of the noise spectrum matrix are
\begin{widetext}
\begin{eqnarray}
N_{11}&=&  \frac{1}{N_g} \left[  \left(\frac{\partial \bar{g_1}(c)}{\partial c} \right)^2 c+\bar{g}_1\right]\\
N_{22}&=&  \frac{1}{N_g} \left[\left(\frac{\partial {f_2}(c, g_1)}{\partial c} {\bigg |}_{g_1 = \bar g_1(c)}\right)^2 c+  \phi_{\rm 21}^2 \frac{\bar{g}_{\rm 1}}{c_{\rm{max}}} + \bar{g}_2\right]\\
N_{33}&=&  \frac{1}{N_g} \left[\left(\frac{\partial {f_3}(c, g_1,g_2)}{\partial c}  {\bigg |}_{\{g_{\rm i} = \bar g_{\rm i}(c)\}}\right)^2 c+ \phi_{\rm 31}^2 \frac{\bar{g}_{\rm 1} }{c_{\rm{max}}} +\phi_{\rm 32}^2 \frac{\bar{g}_2}{ c_{\rm{max}}}+ \bar{g}_3\right]
\end{eqnarray}
Following Eq (\ref{Kfinal}) to find the inverse covariance integral, analogously to the two gene case, requires evaluating the frequency integrals. The general form of these integrals is:
\begin{eqnarray}
I_{ks}&=&\int \frac{d \omega}{2 \pi} \frac{1}{(1+i\omega \tau)^k(1-i\omega \tau)^s}\\ \nonumber
&=&\frac{1}{\tau}\int \frac{d \zeta}{2 \pi} \frac{1}{(1+i\zeta)^k(1-i\zeta)^s}= \frac{1}{\tau} {s+k-2 \choose k-1} 2^{1-s-k},
\end{eqnarray}
 When the dust settles,  the elements of the inverse of the covariance matrix, ${\cal K}^{-1}$ from Eq (\ref{Kfinal}), are:
\begin{eqnarray}
{\cal K}^{-1}_{11}&=&  N_{11}, \\
{\cal K}^{-1}_{12}&=&\frac{1}{2}   N_{11} \phi_{21} ,\\
{\cal K}^{-1}_{13}&=& \frac{1}{4} N_{11} \left[\phi_{21} \phi_{32}+2 \phi_{31} \right], \\
{\cal K}^{-1}_{21}&=&\frac{1}{2}  N_{11} \phi_{21} ,\\
{\cal K}^{-1}_{22}&=& N_{22}+\frac{1}{2}  N_{11}(\phi_{21})^2,\\
{\cal K}^{-1}_{23}&=&\frac{1}{2}  N_{11} \left[(\phi_{21})^2 \phi_{32}+\phi_{31} \phi_{21}\right]+ \frac{1}{2}  N_{22} \phi_{32},\\
{\cal K}^{-1}_{31}&=&\frac{1}{4} N_{11}\left[ \phi_{21} \phi_{32}+2 \phi_{31} \right],\\
{\cal K}^{-1}_{32}&=&\frac{1}{2}  N_{11} \left[(\phi_{21})^2 \phi_{32}+ \phi_{21} \phi_{31}\right]+\frac{1}{2}  N_{22} \phi_{32} ,\\ 
{\cal K}^{-1}_{33}&=&  N_{33}+\frac{1}{2} \left[ N_{22} (\phi_{32})^2+ N_{11}(\phi_{31})^2\right]+N_{11} \phi_{21} \phi_{32} \phi_{31} +\frac{3}{8}N_{11}(\phi_{21} \phi_{32})^2
\end{eqnarray}
\end{widetext}

\begin{acknowledgments}
We thank  J Dubuis, T Gregor, W de Ronde, EF Wieschaus, and especially CG Callan for helpful discussions.  Work at Princeton was supported in part by NSF Grant PHY--0650617, and by NIH Grants P50 GM071508 and R01 GM077599.  GT was supported in part by NSF grants DMR04--25780 and IBN--0344678, and by the Vice Provost for Research at the University of Pennsylvania.  WB also thanks his colleagues at the Center for Studies in Physics and Biology at Rockefeller University for their hospitality during  a portion of this work.
\end{acknowledgments}


\begin{thebibliography}{99}
\bibitem{dale}
There are examples of transcription factors that act on one target as an activator and on another target as a repressor, but it is unusual that this happens with the factor acting alone.  In eukaroytes, where transcription factors often form complexes, an activator without its partner can function as a repressor  \cite{Brivanlou02}.
\bibitem{lifetime_note}
In bacteria, it often is assumed that the dominant term in the decay of protein concentrations is growth and division of the cells, which would make all lifetimes equal.  More generally, we consider all transcription factors to operate under similar constraints, so that all of the network structure comes out of the optimization problem, rather than being imposed by hand.
\bibitem{volume}
We note that, in eukaryotes, the relevant volume for sensing the concentration of transcription factor proteins is the nucleus, however proteins are produced in the cytoplasm and then imported into the nucleus.  Allowing for this partitioning,  $\Omega$ becomes an effective volume, not the literal volume of the cell or nucleus.
\bibitem{unequalK}
We have also studied the general model, with both $K_{\rm on}$ and $K_{\rm off}$ parameters.  The optimum information transmission always occurs at  $L\gg1$, and the details of the optimal solution reproduce the behavior in the simpler limit considered here.
\end{thebibliography}
\end{document}